\documentclass[useAMS,usenatbib,usegraphicx]{mn2e}

\usepackage{graphicx}
\usepackage{amsmath}
\usepackage{multirow}
\usepackage{lscape}
\newcommand{\hbeta}{\hbox{H$\beta$}}
\newcommand{\othree}{\hbox{[O\textsc{iii}] $\lambda\lambda 4959,5007$}}
\newcommand{\othreeb}{\hbox{[O\textsc{iii}] $\lambda 4959 $}}
\newcommand{\othreec}{\hbox{[O\textsc{iii}] $\lambda$5007}}
\newcommand{\halpha}{\hbox{H$\alpha$}}

\newcommand{\ntwo}{\hbox{[N\textsc{ii}]$\lambda\lambda 6548,6583$}}

\def\sun{\hbox{$\odot$}}

\def\lesssim{\mathrel{\hbox{\rlap{\hbox{\lower4pt\hbox{$\sim$}}}\hbox{$<$}}}}
\def\gtrsim{\mathrel{\hbox{\rlap{\hbox{\lower4pt\hbox{$\sim$}}}\hbox{$>$}}}}

\def\arcdeg{\hbox{$^\circ$}}
\def\arcmin{\hbox{$^\prime$}}
\def\ion#1#2{#1$\;${\small\rm\@Roman{#2}}\relax}

\def\ntd#1{\vtop{\footnotesize\hsize=\textwidth\leavevmode#1\hspace*{\fill}}}

\title{Planetary Nebulae towards the Galactic bulge. \\ 
I. [OIII] fluxes}
\author[A.~V. Kovacevic, Q.~A. Parker, G.~H. Jacoby, R. Sharp, B. Miszalski and D.~J. Frew]{Anna~V. Kovacevic$^{1}$\thanks{Correspondence to E-mail: akovac@ics.mq.edu.au}, Quentin~A. Parker$^{1,2}$, George~H. Jacoby$^{3}$, Rob Sharp$^{4}$, 
\newauthor Brent Miszalski$^{5}$ and David~J. Frew$^{1}$\\
$^{1}$Department of Physics, Macquarie University, North Ryde, NSW 2109, Australia\\
$^{2}$Anglo-Australian Observatory, PO Box 296, Epping, NSW 1710, Australia\\
$^{3}$NOAO, 950 North Cherry Ave. Tucson, AZ 85719, USA \\ 
$^{4}$Research School of Astronomy \& Astrophysics, Mount Stromlo Observatory, Cotter Road, Weston ACT 2611, Australia\\
$^{5}$Centre for Astrophysics Research, STRI, University of Hertfordshire, College Lane Campus, Hatfield AL10 9AB, UK
}

\begin{document}

\date{Accepted 2010 MONTH XX. Received 2010 MONTH XX; in original form 2010 MONTH XX}

\pagerange{\pageref{firstpage}--\pageref{lastpage}} \pubyear{2010}

\maketitle

\label{firstpage}

\begin{abstract}
We present $\othreec$ fluxes and angular diameters for 435 Planetary 
Nebulae (PN) in the central $10\arcdeg\times10\arcdeg$
region towards the Galactic bulge. Our sample is taken from the new discoveries of the MASH PN 
surveys as well as previously known PN. This sample accounts for 80~per cent of known PN in this region. 
Fluxes and diameters are measured from narrow-band 
imaging with the MOSAIC-II camera on the 4-m Blanco telescope 
at the Cerro-Tololo Inter-American Observatory. This is the largest ($\sim$60 square degrees), 
uniform $\othreec$ survey of the inner Galactic bulge ever 
undertaken. 104 of the objects have measured 
$\othreec$, $\othreeb$, $\halpha$ or $\hbeta$ fluxes from the literature, which we use 
to undertake a detailed comparison to
demonstrate the integrity of our new fluxes. 
Our independent measurements are in excellent agreement with the very best literature 
sources over two orders of magnitude, while maintaining good consistency over five orders of magnitude. 
The excellent resolution and sensitivity of our data allows not only for a robust set of homogenous PN fluxes, 
but provides greater detail into their intricate, otherwise undetermined $\othreec$ morphologies.  
These new, extensive measurements significantly increase the 
sample of reliable $\othreec$ fluxes for Galactic bulge PN making it a valuable resource 
and a prelude to the construction of our new Galactic bulge Planetary Nebula luminosity function (Paper II).  
\end{abstract}


\section{Introduction}
\label{sec_intro}
Since the realisation that the Planetary Nebula luminosity 
Function (PNLF; Jacoby, Ciardullo \& Ford 1988) can be used 
as a standard candle, it has been firmly established as a reliable extra-galactic distance 
indicator for galaxies out to the Virgo (18.0$\pm$1.2~Mpc; \citealp{2001A&A...375..770F}) 
and Fornax (18.6$\pm$0.6~Mpc; \citealp{1999ApJ...515...29M}) clusters and beyond \citep{JCF90, FJP07, G07,C10}. 
Agreement with traditional population-I distance indicators such as 
Cepheids and SN~Ia \citep{C03} provides us with a rare link with which we can 
compare these independent methods between very different stellar populations 
of different ages and metallicities. However, despite the 
robustness of the PNLF as a distance indicator, the underlying detailed physics responsible 
is not well understood. The strength of the PNLF distance technique resides 
in the fit of the PNLF function having a definitive absolute magnitude bright-end cut-off at 
M$_{\mathrm{[OIII]}}=-4.47^{+0.02}_{-0.03}$ \citep{C02}, 
regardless of galaxy type and age, though there does appear to be a 
small dependence on metallicity \citep{D92}. \citet{C05} outlined this 
fundamental inconsistency. The PN at the bright-end of the PNLF have central star 
luminosities of $\sim$6000~L$_{\sun}$, which requires the central stars to be $>$0.6~M$_{\sun}$ 
\citep{2008ApJ...681..325M,2004A&A...423..995M} and corresponds to a mass on the 
main-sequence of $>$2~M$_{\sun}$ \citep{W00}. 
Stars with such main-sequence masses, and consequent short lifetimes ($\sim$1-2~Gyr), are not expected to 
be present in elliptical galaxies whose populations are 
typically $\sim$10~Gyr old and where there is no evidence for recent star formation. 
However, the bright PN that have evolved from such stars are nonetheless 
detected in these populations.

This conundrum has been a major obstacle in our interpretation 
and understanding of the apparently fixed nature of the bright-end of the PNLF across galaxy types 
for nearly 20~yrs. To address the problem we urgently need to deconstruct 
the PNLF in fine detail. \\

To evaluate the varied characteristics of individual PN constituting 
the PNLF, we require a self-contained system at a known distance whose 
PN population is sufficiently nearby to permit investigation into individual 
PN morphologies, their abundances, expansion velocities, central star temperatures and consequently 
their masses. Such detailed analyses can be achieved in relatively nearby galaxies such as the LMC 
\citep{2010arXiv1002.3410R} and SMC \citep{2002AJ....123..269J}, 
but being metal-poor and comprised of intermediate-age populations they cannot represent valid proxies for an old elliptical galaxy 
and cannot be used to address the problem outlined above. \\

%Detail analyses of the next 
%nearest populations (M31 bulge and M33 bulge) would require significantly large amounts of observing time, 
%and to decipher their morphologies would be infeasible with current imaging technologies.

Fortunately, the bulge of our Galaxy, being relatively nearby ($\sim$8~Kpc), 
has the potential to represent such a system. There is mounting 
evidence that our Galactic bulge formed within a very short time-scale 
\citep{rich05,zoccali06,fulbright07} 
about $\sim10$~Gyr ago 
\citep{ortolani95,feltzing00,Z03}. A rapid 
star formation history in the early universe indicates that early-type 
spiral bulges undergo a comparable formation mechanism 
to elliptical galaxies \citep{peletier99,falcon02}.
We can therefore exploit its proximity and population age as a proxy for an elliptical galaxy amendable to detailed study. 

This work has the potential to determine whether the PNLF bright-end is 
comprised of PN resulting from old, population-II stars which have found some 
peculiar path to enhanced luminosity, e.g. through binarity \citep{C05,M09a}, or if it 
is in fact dominated by younger, higher mass, 
bipolar nebulae mainly of Type-I as defined by \citet{KB94} and \citet{TPP97}. 
These are thought to derive from more massive progenitors which suffer third 
dredge-up and undergo hot-bottom burning, subsequently becoming nitrogen and helium enriched. 
Such detailed analysis of an old stellar population is currently, and for the forseeable future,  
impossible to do anywhere else. The next nearest old population is the bulge of 
M31 and in order to obtain spectra of comparable quality/depth to what exists for Galactic bulge PN 
requires instrumentation beyond what is currently available. Hence, this work will provide 
a significantly improved understanding 
of the PN population in all elliptical galaxies as well as in spiral bulges. \\

Our ability to study such samples has been recently enhanced 
thanks to the highly sensitive AAO/UKST SuperCosmos 
$\halpha$ Survey of the southern Galactic plane (SHS; \cite{P05}) which 
enabled doubling the number of known bulge PN in this region as reported in the 
Macquarie/AAO/Strasbourg $\halpha$ PN catalogues of \citet{P06} and \citet{M08}, dubbed MASH and MASH-II.  
This combined sample is more representative of the bulge PN population spanning a wider 
evolutionary range and allows us here to construct a new $\othreec$ 
Galactic bulge PNLF of unprecedented coverage. 
This will enable detailed study of the various sub-sets of 
PN (whether Type-I chemically or of certain morphologies) 
within the PNLF, whether they are confined to certain regions within the PNLF and, importantly, 
whether one population sub-set is solely responsible for constituting the bright-end. These 
analyses will be presented in \citet{K10b}, hereafter Paper II, while this paper describes the process for 
defining robust $\othreec$ flux measurements. \\

\section{Observations}
\label{obs}

\begin{figure*}
\centering
\includegraphics[scale=0.6]{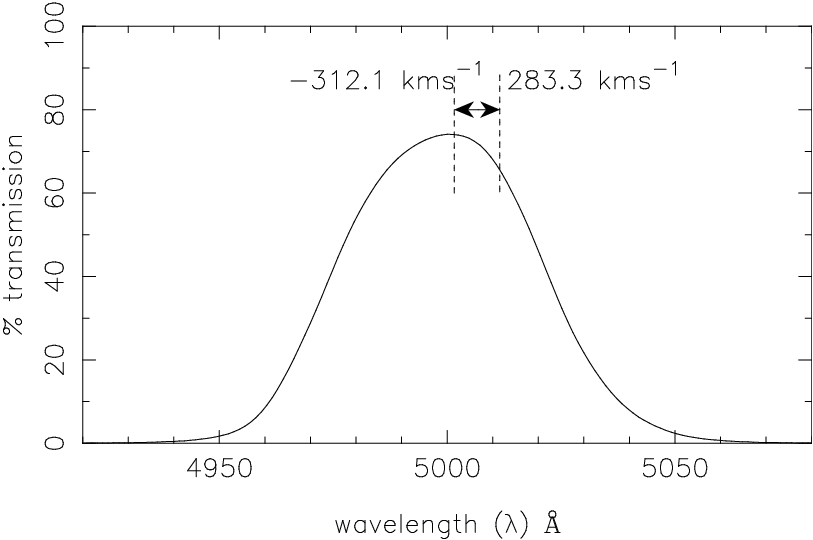}
\includegraphics[scale=0.6]{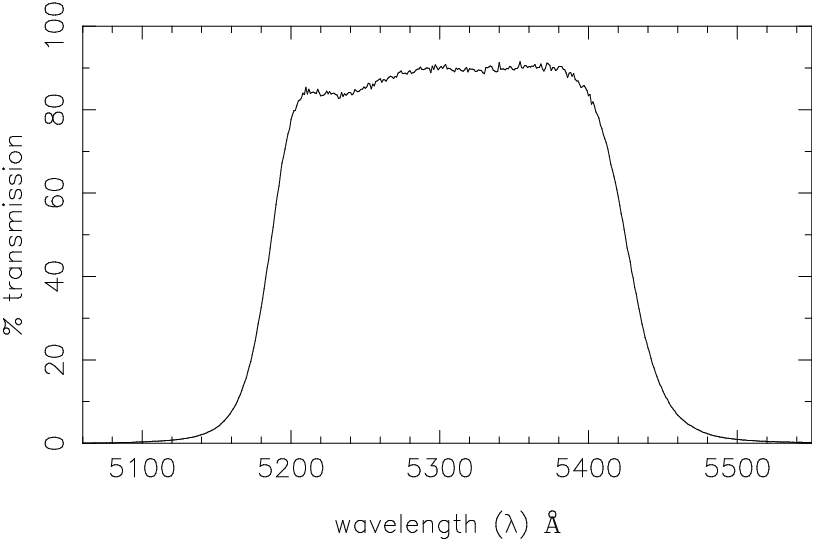}
\caption{The f/2.9 transmission curve of the $\othreec$ (left) and off-band $\othreec$ filter (right) used. 
The wavelength shift of the $\othreec$ line experienced by the PN with largest positive and negative 
velocities are shown here. The filter transmission experienced by these two PN is 74.07~per cent and 64.71~per cent, respectively.}
\label{fig:filterOIII}
\end{figure*}

We observed previously known and MASH PN in a $10\arcdeg\times10\arcdeg$ region toward the 
Galactic bulge using the Cerro-Tololo 
Inter-American Observatory (CTIO) 4-m Blanco telescope in Chile with the 
MOSAIC-II CCD Imager \citep{M98}.
This instrument provides a field of view (FoV) of 
$\sim$36$\arcmin\times36\arcmin$, using a mosaic of 
eight SITe $2048\times4096$ pixel CCDs, at a scale of 
$\sim$0.27~arcsec~pixel$^{-1}$. A total of $\sim$95~hours of photometric 
imaging was conducted, spread over two observing runs: 
6 nights throughout 9th -- 14th June 2008 and 5 nights from 27th June -- 2nd July 2009. 
The weather was consistently photometric throughout the first run with a 
modest seeing range from $0.7 - 1.0$~arcsec, with only 1.5 hours lost to technical and software faults. 
The weather was significantly more variable throughout the 2009 run, 
resulting in a loss of 2.5 nights.
In photometric conditions, the seeing varied from $1.0 - 2.0$~arcsec, 
with extremes of 2.5~arcsec. Exposure times comprised of 3$\times$400~second in the $\othreec$ band, 
an additional 60~second exposure for those fields containing 
bright PN to avoid saturation, and 3$\times$90~second exposures in the 
broader, $\othreec$-off band filter. In the 2009 run we modified the number of off-band 
exposures to 2 in order to cover a wider area while still reaching an adequate 
depth. Filter details are given in Table \ref{tab:filters}, 
while the transmission curves of the $\othreec$ on- and off-band filters are shown in 
Fig. \ref{fig:filterOIII}. The $\othreec$ transmission 
curve was measured by NOAO with a simulated f/2.9 beam for the CTIO 4-m\footnote{Filter curves can be found 
at http://www.ctio.noao.edu/mosaic/manual/index.html}, 
whilst the off-band transmission 
curve was measured with a simulated f/3.1 beam for the KPNO 4-m, so should 
be $\sim$3\AA\ bluer when used on the CTIO 4-m.   
In addition, spectrophotometric standard stars were observed at the beginning, 
middle and end of each night to allow for flux calibration of the fields.\\

\begin{table*}
 \caption{Filters used in the CTIO observing runs.}
 \begin{tabular}{ l| l || l || l | }\hline
 \hline 
 Filter & NOAO code & $\lambda_{central}$ (\AA) & FWHM (\AA) \\
\hline 
$\othreec$ & c6014 & 5000 & 50 \\
$\othreec$+30nm & k1015 & 5320 & 240 \\
$\halpha$+$\ntwo$ & c6009 & 6563 & 80 \\
$\halpha$+8nm &  c6011 & 6650 & 80 \\
% V Harris c6026 & 5370 & 940\\ 
\hline      
\end{tabular}
\label{tab:filters}  
\end{table*}

As part of general follow-up, we also have obtained 100~per cent spectroscopy for all MASH PN in the bulge sample \citep{P06,M08}, including at 
high resolution for radial velocity determination. Although the spectra were primarily taken for identification and radial velocity purposes, they can give 
initial estimates of the Balmer decrement in many cases. These can be used to provide preliminary dereddened fluxes (see section \ref{sec:deredden}).

\subsection{Strategy}
\begin{figure*}
\centering
\includegraphics[scale=0.9]{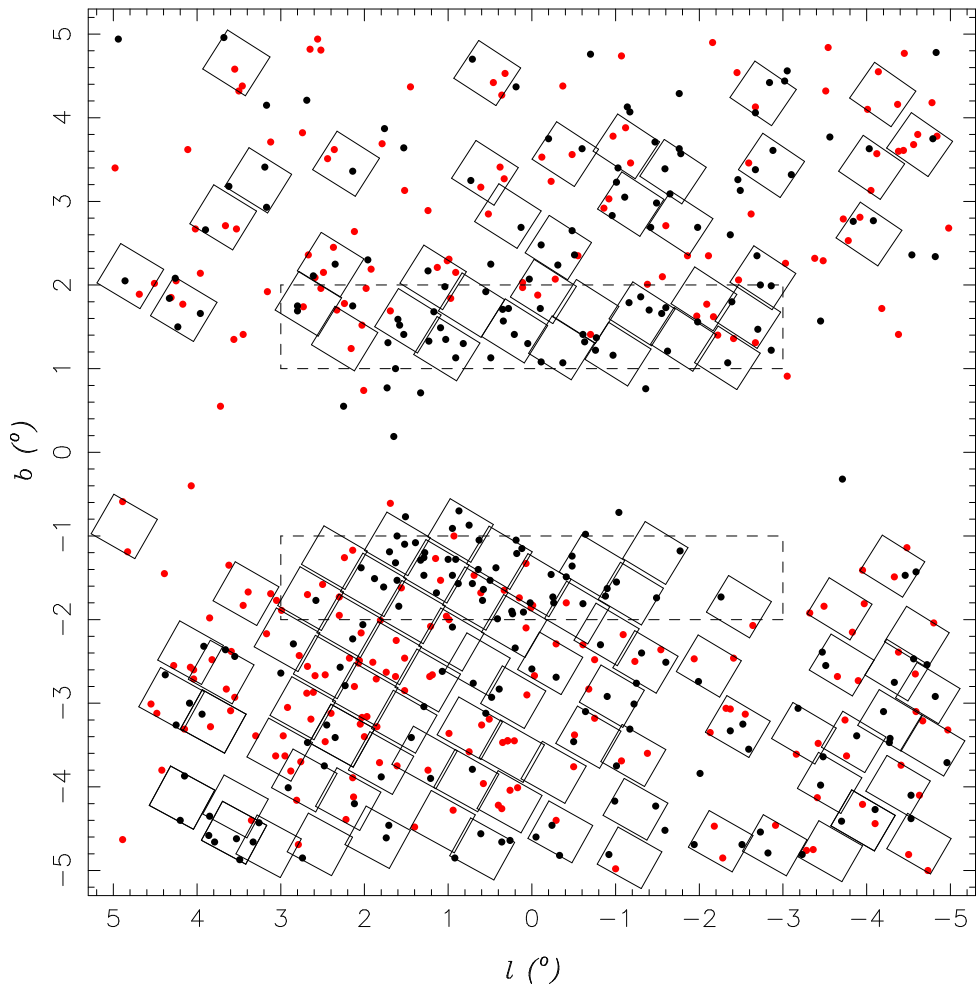}
\caption{Placement of the fields in the direction of the Galactic bulge. The dots illustrate the 
distribution of PN, the red circles being MASH-I and MASH-II PN while 
the black circles represent previously known PN. The squares 
are the observed MOSAIC-II fields. }
\label{fig:bb}
\end{figure*}

The large field-of-view ($36\arcmin\times36\arcmin$) of the MOSAIC-II camera allowed for the 
observation of $\sim$4 PN per field. Placement of fields were 
chosen manually using the Aladin software \citep{B00}, with a MOSAIC-II template especially 
provided. For regions where the PN number 
density was significantly higher, 
the field placement was arranged in a tiled 
manner, allowing an overlap of $\sim2\arcmin$ between fields. 
This applied to the $358\arcdeg\leq l\leq2\arcdeg$ and $-1\arcdeg\leq |b| \leq-3\arcdeg$ region. 
For the remaining area, field placement was chosen such that the number of fields 
was minimised whilst ensuring observation of a maximum number of PN. Fig. \ref{fig:bb} 
illustrates the coverage of survey fields. \\

Approximately 150 pointings were needed to obtain 
$\othreec$ images for all PN in this region. We were able to observe 124 fields, 
covering 80~per cent of all MASH and previously known PN in the 10\arcdeg$\times$10\arcdeg region towards the 
Galactic bulge. Table \ref{tab:fields} lists the identification assigned (column 1) of the fields 
observed, and their central co-ordinates in RA (column 2) and declination (column 3) with the 
exposure times of images (column 4). 
To avoid incompleteness in 
specific regions of the bulge due to potential adverse 
weather conditions during either run, we decided to focus each run on observations 
of either the north or south of the bulge. 
As the southern half of the bulge is the most densely populated with PN 
(suffering lower extinction), we largely dedicated our 
2008 run to observing PN in this half, and observed northern PN on our 
2009 run. As the northern bulge rises $\sim$30~mins before the southern half, we 
were also able to observe one northern bulge 
field in the 2008 run, and an extra southern bulge field in the 2009 run. 
The majority of fields were observed at an airmass of $\leq$~2.

\subsection{Resolution and uniformity}
The uniformity of this $\othreec$ survey not only provides accurate fluxes for the largest sample of bulge PN to date, 
but the high resolution of the MOSAIC-II camera allows for precise angular 
diameter measurements (see section \ref{sec:diameters}) and a greater insight into the 
details of the morphologies of these predominantly compact 
PN than previously available. Better, well-defined classifications of PN morphologies can provide clues as to 
their modes of evolution \citep{M09a,M09b,K09}, and can therefore help isolate sub-sets of the PN population 
when analysing their placement on the PNLF. This issue will be addressed further in Paper II in this series.

\section{Reductions}
\label{reduc}
The data were reduced using the Cambridge Astronomical Software Unit (CASU) pipeline/toolkit \citep{IL01}. 
Although this toolkit was originally written for 
the Wide-Field Imager on the Isaac-Newton Telescope (INT), additional 
software was available to extract systematic errors specific to MOSAIC-II data. \\

Photometry was performed on the cross-talk corrected (e.g. see \citealt{freyhammer01}), bias-subtracted, 
flat-fielded, astrometrically calibrated, continuum-subtracted images. 
Multiple images of the same field were taken to both avoid CCD saturation 
issues and then stacked to eliminate cosmic-rays present in the images. 
The USNO-A2.0 catalogue was 
used to derive the astrometric fit. 
Removal of the stellar and background continuum for each field was achieved by comparing intergrated 
counts of $\sim$50 stars in the off-band 
image to the $\othreec$ image. This determined a flux scaling factor which, once applied to the off-band image, 
ensured approximately equivalent 
depths for continuum point sources between bands.
The off-band image was then brought into alignment 
to the on-band image by applying the XREGISTER routine 
in IRAF. This removed any non-integer pixel shifts in the x- and y-direction, 
and accounted for any rotational and stretching effects.
The off-band image was then subtracted from the on-band image.
This procedure was successful in the effective removal of stars for 
the majority of bulge fields, but not so 
for fields that exhibited very high stellar number densities.  \\

Fluxes were then measured using the aperture photometry utility within 
GAIA\footnote{http://www.starlink.ac.uk/gaia}, with the average sky determined 
via a clipped mean, in order 
to exclude deviant pixels from the calculation. Aperture sizes were chosen 
to be as large as possible without inclusion of any badly-subtracted stars. The flux contributions of such 
artefacts were individually taken into account when they directly overlaid the nebula. Such instances
were of common occurence in extended PN ($\gtrsim20$~arcsec). Fluxes were only measured for the 
bright PN where the maximum pixel value was $\lesssim30000$ counts, as the 
MOSAIC-II CCDs are documented\footnote{http://www.noao.edu/ctio/mosaic/} to 
show non-linear characteristics beyond $\sim35000$ counts per pixel. \\

\subsection{Flux calibration}
The $\othreec$ fluxes were calibrated in units of \linebreak erg s$^{-1}$ cm$^{-2}$  relative to the following spectrophotometric standards: 
EG 274, LTT 9491, LDS 749B, LTT 4816, GD 108 and LTT 7987, 
whose spectra are highly sampled and are well-documented in the literature \citep{B96, H92}. 
The flux calibration procedure followed that of \citet{JAQ87}. 
The integral under the filter curve was multiplied by the flux at 5006.8\AA~ (the rest wavelength of the 
$\othreec$ line) and  
divided by the count rate of the standard star to achieve a system sensitivity in 
ergs s$^{-1}$ cm$^{-2}$ count$^{-1}$. 
The global counts from each PN were then converted into 
a count rate by dividing by the length of exposure time and corrected for 
atmospheric extinction which was taken to be 0.17 mag/airmass \citep{H92}. \\
The radial velocities of 
the PN were utilised to determine the shift of the 
$\othreec$ line, and consequently to determine the transmission of the filter at that wavelength. 
This is especially 
important for bulge PN as they can have relatively high velocities ($\sim\pm300$ km/s). 
Additionally, the shape of the filter 
displayed a gaussian-like profile, so the transmission through the filter for the PN with the most negative velocity 
(74.07~per cent) was significantly different than that from the PN with the most positive velocity (64.71~per cent). 
These fluxes would have been subjected to an additional error of $\pm^{0.05}_{0.02}$ in the 
-$\log$F had this not been taken into account. 
The bulk of non-MASH PN velocities were taken from 
\citet{D98} and the remainder were taken from \citet{M10} (hereafter D98 and M11, 
respectively). The velocity measurements from these two sources combined account for all but 34 of the 
objects measured. To ensure consistency, velocity measurements from other sources such 
as low dispersion object prism spectra were not considered. Fluxes were then calculated 
by converting a count rates to a flux via multiplication by the system sensitivity.

\section{{\hbox{[O\textsc{iii}] $\lambda$5007}} fluxes}
\label{results}
We present both our observed $\othreec$ global fluxes and preliminary dereddened fluxes, and angular diameters for 
435 PN in the $10\arcdeg\times10\arcdeg$
region toward the Galactic bulge in Table \ref{tab:fluxes}, supplemented by a further 6 outside of 
this region, bringing the total observed to 441. 

In this central bulge $10\arcdeg\times10\arcdeg$ sector, the literature 
contains directly measured $\othreec$ fluxes for 48 PN, 
therefore this paper is contributing new fluxes for 387 PN in this region.  
Their IAU PN G and common names are listed in the first 
two columns of Table \ref{tab:fluxes}, the observed $\othreec$ flux 
measured directly from the CCD image is presented in column (3) (in erg cm$^{-2}$s$^{-1}$) with their 
estimated errors in column (4), while their preliminary dereddened $\othreec$ flux is listed in 
column (5). Their major and minor axes 
(in arcsec) are listed in columns (6) and (7). 
We have tabulated five other known PN that were observed in the peripheral region and 
the recently discovered open cluster PN \citep{P10} in Table \ref{tab:extra}. 
%All fluxes presented are uncorrected for interstellar absorption. 
PN that display no observable 
flux are tagged as non-detections where they are either
too cool to produce $\othreec$ emission or too heavily reddened to meet
our detection/sensitivity limit. Fluxes were dereddened for those PN with suitably 
calibrated DBS or AAOmega spectroscopy. This was available for $\sim60$~per cent of those 
PN with observed fluxes.

\subsection{Error analysis} 
\label{sec:errors}

Total errors in our observed flux measurements were derived for each PN. These took into account the 
errors due to the strength of the signal, the velocities of the PN, the 
transmission profile of the filter curve, and those due to variation in the sensitivity factor 
calculated from the spectrophotometric standard stars. \\

%{\bf{Photon statistics}}\\
The error due to the signal strength was determined using the full CCD photometric S/N 
calculation as given in equation \eqref{eq:signoise} \citep{MH05}. 

\begin{equation}
\label{eq:signoise}
\frac{S}{N} = \frac{N_\ast\times G}{\sqrt{N_\ast\times G + n_{pix}(1+n_{pix}/n_B)(N_{S}\times G + N_{D} + N^{2}_{R} + G^{2}\sigma^{2}_{f})}}
\end{equation}

Where N$_\ast$ is the sky-subtracted, summed counts within the aperture, n$_{pix}$ is the number 
of pixels in the area of the aperture, N$_{S}$ is the average counts of the background sky per 
pixel, n$_{B}$ is the number of pixels that have been used to determine N$_{S}$, N$_{D}$ is the 
number of dark current electrons per pixel, N$_{R}$ is the number of electrons per pixel due to 
read-out noise, G (e$^{-}$/ADU) is the gain of the CCD, and $\sigma^{2}_{f}$ is the estimate of 
the 1 sigma error introduced within the A/D converter. The denominator comprises all of the 
relevant factors that contribute to the noise for an ideal system. \\

The conversion from ADUs to electron counts was achieved via multiplication of the pixel counts 
by the gain of the CCD, which is 2~e$^{-}$/ADU for the SITe CCDs on the 
MOSAIC-II camera. They have a read-out noise of 6-8e$^{-}$ and a dark-current 
of $\lesssim$ 2~e$^{-}$pixel$^{-1}$hour$^{-1}$. As the areas of the apertures may be considerable 
in the case of extended PN, it is important that we retain a constant ratio between the area of the 
nebula aperture and the total area of the sky aperture used, $(1+\frac{n_{pix}}{n_{B}})$.

\begin{figure}
\begin{center}
\includegraphics[scale=0.35]{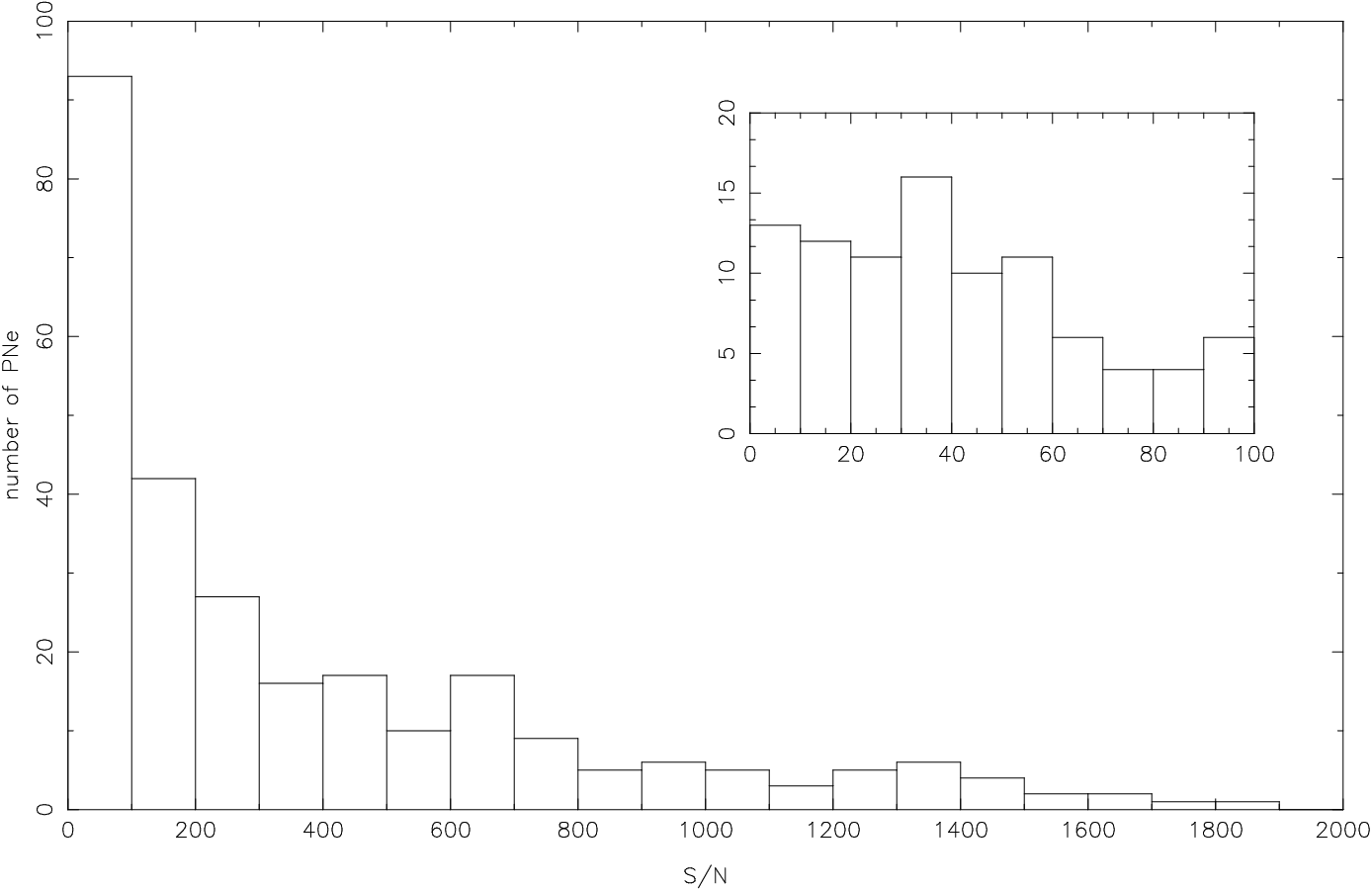}

\caption{Distribution of S/N measurements. The inset graph magnifies the distribution 
0 $<$ S/N $\leq$100. Only 13 PN have S/N less than 10. }
\label{fig:SNdistribution}
\end{center}
\end{figure}

Fig. \ref{fig:SNdistribution} displays the distribution of S/N measurements for all PN we 
observed. This only shows those measurements whose S/N$\leq2000$. There 
were 8 measurements of bright PN which exceeded this. The S/N distribution within the $0<$S/N$\leq100$ range 
is relatively flat, with only 13 PN, or 0.3~per cent of the population, having low S/N $<10$.\\

The magnitude of flux error due to the uncertainty in the velocities were determined 
from the errors published in the D98 and M10 sources. 
At the time of communication, small errors (typically $\sim$1--5~kms$^{-1}$) were 
only available for half of the M10 sample (called M1), so we calculated velocity errors for the  
remaining PN (sample M2) via calculation of the standard deviation $\sigma$ of 
the difference between 
the M1 and M2 samples. This resulted in an error for all M2 velocities of $\pm9.5$~kms$^{-1}$. 
For 33 PN where velocity information did not 
exist, we adopted a value of the average velocity of {\it{all}} other PN 
in the $10\arcdeg\times10\arcdeg$ which have velocities from D98 or M10, 
and adopt the standard deviation 
of the distribution as an estimate of the error. 

Errors 
calculated for the system sensitivity were determined by taking the standard deviation of 
all individual measurements from each standard star. The errors 
due to the sampling of the filter profile were calculated 
from the 1.2~\AA~incremented 
transmission curve determined by NOAO (see fig. \ref{fig:filterOIII}). There was also 
the possibility that the shape of the transmission curve had altered due to filter ageing. The possible 
influence of this effect is assessed in section \ref{sec:duplicate}.

The only other source of error we were not able to account for was 
the component dependent on which regions of sky were chosen. This was a function of how 
cleanly the sky had been subtracted, and the size of the sky regions selected.

After the off-band image had been scaled to and subtracted from the on-band image, 
the colour differences in the stars would often leave 
positive and negative residuals. Differences in seeing and focus between on- and off-band fields 
also resulted in imperfect alignment which accentuated any image residuals. This 
results in a broader range of sky values contributing to the sky background 
determination. Therefore, for PN in very crowded fields, there was an increased 
likelihood that slight deviations in non-perfect subtraction would 
amount to a larger error in the sky background. This source of error would 
dominate for PN with large angular diameters, 
where it was 
necessary to extract equivalent areas of sky, as it was likely that the sky areas chosen 
would include more artefacts.  

Therefore, for PN in regions of high stellar number density, and particularly of 
large angular diameter, there would be larger errors 
associated with the average value of the background sky, and subsequently in the measured 
flux. This is potentially the dominant source of error, and is discussed further in section 
\ref{sec:duplicate} below. \\

\subsection{Comparison of duplicate observations of PN}
\label{sec:duplicate}
\begin{figure}
\begin{center}
\includegraphics[scale=0.6]{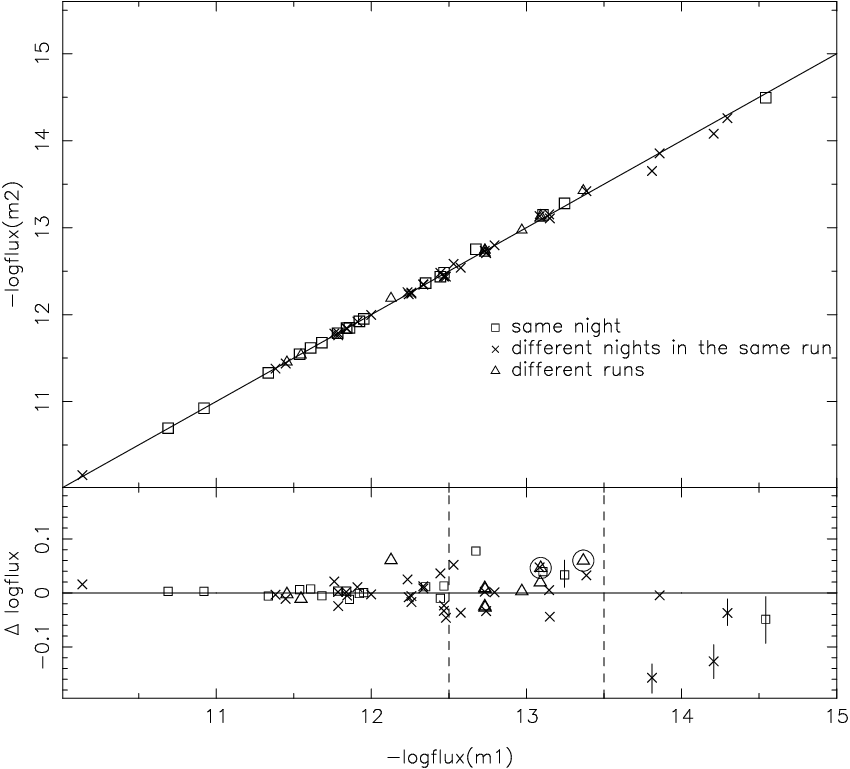}

\caption{Comparison of observed $\othreec$ fluxes obtained for repeat measurements of PN. 
Shapes of the data points signify whether the measurements were from fields observed 
in the same night, for a different night during the same run, 
or different runs.  
Errors between measurements have been plotted for individual points when 
they exceed $\pm0.025$, otherwise the error is indicated by size of the point. The dashed lines indicate the 
boundary between regions encompassing similar $\delta\log$F measurements. }
\label{fig:errorsMulpMeas}
\end{center}
\end{figure}

During this project we obtained 62 duplicate observations of PN, 
whether on the same night (a), different nights 
within the same run (b), or different runs (c). To ensure appropriate error assignments 
determined from the previous section, 
differences between multiple $\log$F measurements were calculated and 
are presented in Fig. \ref{fig:errorsMulpMeas}. 

Circled points from the lower panel of Fig. \ref{fig:errorsMulpMeas} are excluded from the final comparison as for these two points 
the seeing was significantly higher for one epoch, which led to 
large discrepancies between compared fluxes. 
Indeed, fluxes of the PN observed in fields of poorer 
seeing were found to be consistently fainter. This was the case 
for these two PN, where differences between their measurements were 0.060 and 0.046. Both data 
points were constituents of group (c). 
Careful comparison of the seeing conditions in all fields revealed 
that there were only two instances where the 
fields had significantly worse seeing ($\sim$~2.5~arcsec as 
opposed to $\sim$~0.7~arcsec seeing). 
Please note that PN with 
fluxes measured on either of these fields will 
be published with an error of $\pm0.063$ in the $\log$F. 
However, measured fluxes were discarded if the PN has a repeat flux measurement in 
another field with considerably better seeing conditions.  \\

By comparing all duplicate measurements, there appears to be trends fitting 
three distinct regions in the flux error distribution. These are divided via the dashed lines in 
Fig. \ref{fig:errorsMulpMeas}, with their sample sizes and standard deviation of all points in each region 
presented in Table 2. 
To ensure we have included 
all small systematic deviations in our error assignment, we designate 
these standard deviations as the error 
in our $\log$F measurements, apart from those PN where their error from low S/N dominates. 
Here, their individual calculated errors are quoted. \\

\begin{table}
\centering
  \caption{Three regions in the $\Delta\log$ flux distribution have been identified.
The range in $-\log$F, standard deviation and number of data points in the group are tabulated. }
\begin{tabular}{rclll}
\hline
\multicolumn{3}{|c|}{$-\log$F range} & $\sigma$ & N \\ 
\hline
        & $-\log$F & $\leq12.5$ & 0.02 & 36 \\
12.5 $<$ & $-\log$F & $\leq13.5$ & 0.03 & 19 \\
13.5 $<$ & $-\log$F & $\leq15$   & 0.09 & 5 \\
\hline
\end{tabular}
   \label{tab:errors}
\end{table}

%Therefore, for PN in regions of high stellar number density, and particularly of 
%large angular diameter, there would be larger errors 
%associated with the average value of the background sky, and subsequently in the measured 
%flux. This is potentially the dominant source of error, and is discussed further in section

\begin{figure}
\begin{center}
\includegraphics[scale=0.6]{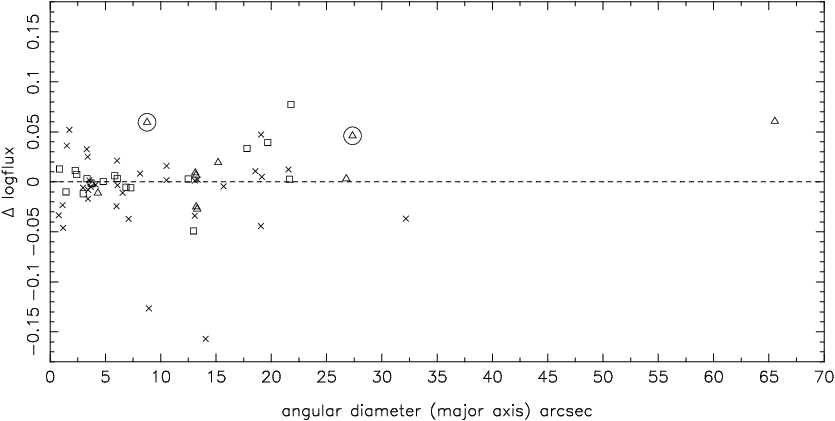}

\caption{PN angular diameters are compared to the difference in flux between 
repeat measurements. The symbol shapes are the same as in Fig. 5. The difference 
between duplicate flux measurements is seen to be 
independent of angular diameter of the PN. }
\label{fig:diameter}
\end{center}
\end{figure}

The uncertainty in the sky subtraction is expected to increase as a function 
of the stellar number density in the field in which the PN is situated, and the angular diameter of the PN 
(since it was necessary to extract an equivalent area of sky), as described towards the end 
of section \ref{sec:errors}. 
The difference in flux between two measurements is therefore proportional 
to the stellar number density of the environment and the angular diameter of the PN. 
Fig. \ref{fig:diameter}, however, shows no angular diameter dependency exists with $\Delta\log$F. Extended PN are not 
observed to have significantly 
larger differences between duplicate flux measurements than smaller, more compact PN. \\

An error 
that we were unable to account for was the extent to which the transmission curve 
of the filter may have altered due to ageing since the initial measurement of the filter transmission curve. 
Ageing effects slow down quickly after the filter is made, 
with filters of this type becoming stable after a few years. The filters used here are $>$12 years old, so 
we do not expect a shift in the filter transmission curve to occur in our one year run-to-run separation between our measurements. 
%{\bf Note our initial one year 
%run-to-run separation is considered too short to reveal any filter ageing effects. This comparison reveled negligible effects anyway. } 

\begin{figure}
\begin{center}
\includegraphics[scale=0.6]{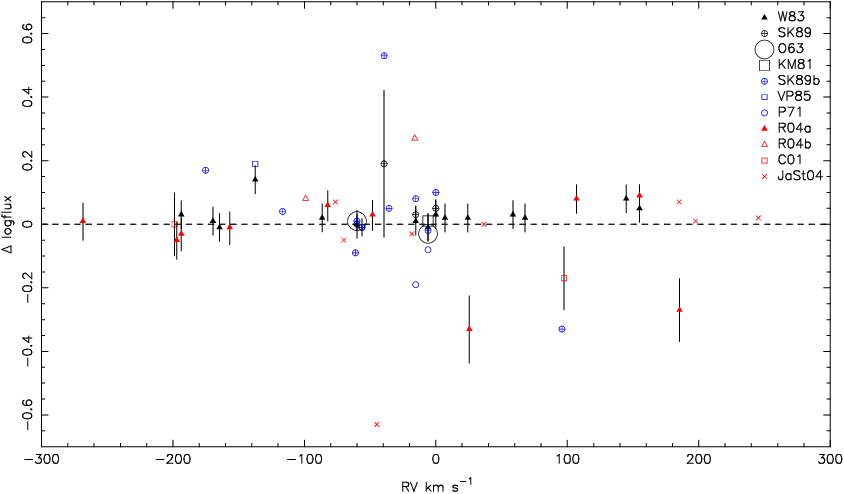}

\caption{ PN radial velocities are compared to the difference between the flux measurement published here and 
that collated from the literature. Black markers indicate where the literature flux 
was measured from $\othreec$ media, whilst red and blue represent those fluxes that have been transformed 
to an $\othreec$ flux from a $\halpha$ or $\hbeta$ flux, respectively. The shape of the data-points signify 
which literature source the comparison has been drawn from. Whilst considering comparisons against 
only the most reliable sources, no dependency on $\Delta\log$flux with PN radial velocity is observed. 
This infers a negligible error contribution from any alteration 
in the filter transmission curve 
due to ageing, which would otherwise accentuate $\Delta\log$F with PN velocity. }
\label{fig:RV}
\end{center}
\end{figure}

%A blue-ward shift of the filter transmission would result in an over-estimation of the 
%The gaussian profile of the transmission curve indicates that the error contribution {\bf (due to the 
%gradient of the curve at that wavelength)} is smaller 
%for PNs with $\othreec$ lines shifted to the blue (due to a negative 
%velocity), and greater for PNs with $\othreec$ lines shifted to the red (due to a 
%positive velocity). 
%If the error in the transmission curve dominated above all others, we would expect $\Delta\log$F to increase with 
%positively increasing PN radial velocity. {\bf Comparison of duplicate observations for PN 
%shows that there is no apparent trend with $\Delta\log$F as a function of radial velocity. We can therefore 
%conclude that the error contribution due to the transmission curve is not a prominent source of error.}
Instead, a significant shift of the transmission profile 
due to filter ageing would result in a more substantial deviation of our fluxes when 
compared to those from the literature, as a function of radial velocity. We refer the reader to 
section \ref{compLit} 
for a full description of the literature sources that we have compared our data to. 
Fig. \ref{fig:RV} plots the difference between fluxes published here and those collated from the 
literature, $\Delta\log$F, against the radial velocity of individual PN. The black data-points represent PN 
where comparison is drawn against an $\othreec$ flux from the literature, and are considered 
to be the most reliable. Red and blue data-points are those where an $\othreec$ flux has been calculated 
from either a $\halpha$ or $\hbeta$ flux combined with a relevant line ratio. 
Whilst considering the most reliable points, it is seen that there is no inclination 
for $\Delta\log$F to bias towards a particular radial velocity. We therefore conclude that potential 
errors due to the ageing of the filter transmission profile are negligible. 

\subsection{Comparison with literature fluxes} 
\label{compLit}
To determine any systematic offset our data may be subject to, we compared our observed 
$\othreec$ fluxes to those PN with published fluxes from the 
literature. Of the PN we observed, 104 had previously published $\othreec$, $\othreeb$, 
$\halpha$ or $\hbeta$ fluxes with a $\othreec/\halpha$ or $\othreec/\hbeta$ line 
ratio where applicable. 
Significant contributions in this region have been made by 
\citet{W83}, \citet{KFL88}, \citet{SK89}, \citet{ASTR91}, \citet{R04} and \citet{JV04} 
hereafter W83, KFL88, SK89, ASTR91, R04 and JaSt04 respectively, 
while fewer have been collated from 
\citet{O63}, \citet{KM81}, \citet{C01}, \citet{P71} and \citet{VP85} hereafter O63, KM81, C01, P71 and 
VP85 respectively. Absolute fluxes from the literature have been measured using 
either photoelectric narrow-band 
interference filter photometry (e.g. W83, SK89), 
CCD photometry (e.g. C01, R04) or from photographic plates (KFL88). Photographically derived fluxes are 
unfortunately subject to large errors, and are occasionally completely erroneous. Fluxes taken from the literature 
have only been included if their errors are quoted as being 
$\leq$0.3~dex. This excludes a substantial fraction of flux measurements of PN with angular 
diameters $\gtrsim$~7~arcsec from the ASTR91 sample. They publish global $\hbeta$ fluxes as measured 
for PN with angular diameters smaller than the slit aperture, 
whereas fluxes for larger objects were calculated via 
multiplication by the geometric ratio of the size of the aperture and diameter of the 
PN. This led to an increased error for PN with angular diameters larger than the slit aperture. \\

It was necessary to place all of the older data onto the modern photometric scale so their 
fluxes reflect the most recent Vega calibration. This requires a correction of 
$-0.02$ for all sources apart from W83, which require 
an offset of $-0.03$, while SK89 and ASTR91 need no such correction. 
Offsets are those derived from the recalibrated nebular fluxes as 
described in \citet{SK82}, and are the same offsets implemented by \citet{CKS92}.

We divide the literature sources into three groups, those which have either a direct $\othreec$ or 
$\othreeb$ flux, which we shall consider to be the most reliable, those where an $\othreec$ flux has been 
transformed via combination of spectroscopy and $\hbeta$ or $\halpha$ photometry, 
and lastly those whose $\othreec$ fluxes were purely derived spectroscopically, e.g. via multiplication 
of an aperture correction, which we shall consider to be the least reliable. \\ 

\begin{figure}
\begin{center}
\includegraphics[scale=0.6]{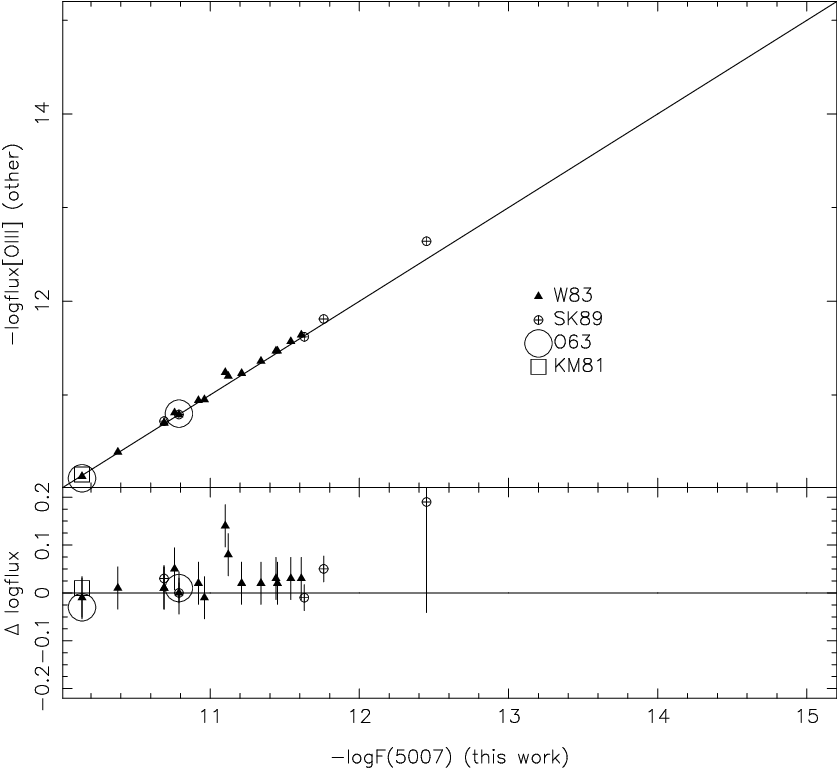}

\caption{Our measured $\othreec$ fluxes compared against other directly measured $\othreec$ 
fluxes compiled from the literature. The 
triangles are data from W83, crossed circles are those from SK89, 
the two larger circles are data from O63 and the single square is from KM81. }
\label{fig:fluxComp}
\end{center}
\end{figure}

O63, KM81 and W83 provide direct measurement of $\othreec$ integrated PN
fluxes, while SK89 have published $\othreeb$ fluxes, 
which are converted into an $\othreec$ flux 
via multiplication by the theoretical value for the $\othreec/\othreeb$ line ratio of 2.98 as given in 
\citet{SZ00}. This sample is illustrated in Fig. \ref{fig:fluxComp}. Note that 
there may be multiple data points for one object, as the brighter objects usually have published fluxes from more 
than one source. This aims to provide a comparison of integrity between literature sources in addition to the 
comparison to our own. It can be seen that the majority of fluxes agree 
with each other to $\pm0.04$ in the log. Considering that the errors from W83 are quoted as $\pm0.04$, 
O63 are $\pm0.01$, KM81 are $\pm0.01$ and those from SK89 are $\pm0.02$ and greater, 
our data are accurate to within the errors of these previously published sources. The SK89 
data point at $\sim$~-12.5 has a larger quoted error of $\pm0.23$, so is still in agreement.

\begin{figure}
\begin{center}
\includegraphics[scale=0.6]{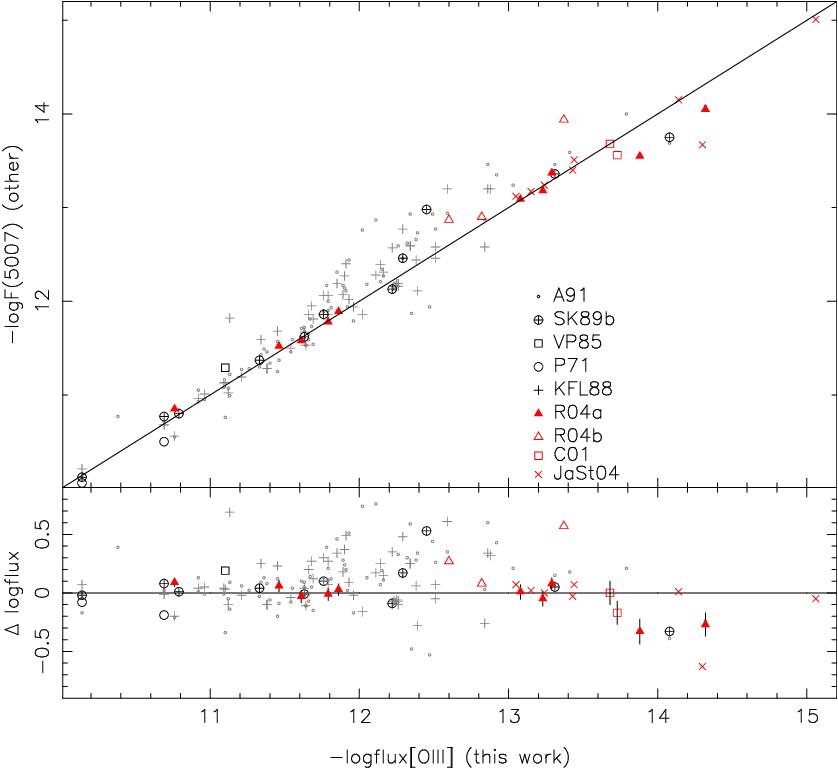}

\caption{A comparison between our $\othreec$ fluxes to boot-strapped $\othreec$ fluxes from the literature. 
Red data-points signify those that have been transformed from a $\halpha$ flux, whereas those in black are from 
a $\hbeta$ flux. The R04 data has been grouped into those where errors exist for both $\halpha$ flux 
and spectroscopic line ratio (filled triangles), and those where they don't (open tiangles). The 
JaSt04 data point with $\Delta\log$F of $\sim$-0.6 is not typical due to the location of the PN (see text).   }
\label{fig:fluxCompBS}
\end{center}
\end{figure}

A secondary, slightly less reliable comparison was made with $\halpha$ (JaSt04, R04, C01) and 
$\hbeta$ (SK89, P71, VP85) fluxes measured from photometry. Their transformation to a 
$\othreec$ flux was calculated via multiplication of their 
$\othreec/\halpha$ or $\othreec/\hbeta$ ratios. 
Spectroscopic line ratios were preferentially taken from data published by \citet{CMKAS00}, 
or \citet{G09} and ASTR91 in the absence of these. 
C01 published line ratios as well as fluxes, so their own line ratios were used here. Unpublished 
spectroscopy \citep{V03} 
for the JaSt04 PN were used to boot-strap their $\halpha$ fluxes to an $\othreec$ flux. 
The comparison of these transmogrified fluxes against 
our measurements are illustrated in Fig. \ref{fig:fluxCompBS}. Note that the 
$\Delta\log$F scale is significantly larger than that in Fig. \ref{fig:fluxComp} . 

The third, least reliable comparison was made where $\hbeta$ fluxes had been derived 
spectroscopically (ASTR91 and KFL88). These fluxes are subject to significant additional error due to the aperture 
corrections made, and therefore can be considered less reliable than the 
previous comparisons. Due to this we include their comparison as fainter data points in Fig. 
\ref{fig:fluxCompBS}. Note that the $\othreec$ fluxes 
included from ASTR91 are those with $\hbeta$ fluxes provided by the authors, not the 
averaged value compiled from previous literature sources. \\

%The aperture corrections made 
%introduce additional error for fluxes within these datasets, and are therefore relative to those in previous comparisons, and can therefore be considered as a lesser 

%In these situations, a $\hbeta$ flux is calculated by consideration of the difference in geometric 
%size of the aperture and angular extent of the PN, and assuming that the  
%from spectrophotometry with the aperture correction for the angular extent of the PN. If the $\hbeta$ flux 
%density is homogenous across the extent of the nebula, the aperture correction should introduce no 
%additional error in the global flux measurement. However, this is rarely the case.

%Due to significantly larger errors inherent to 
%these lower quality datasets, 

As noted by W83, the $\hbeta$ fluxes 
of P71 depend strongly on the $\othreec/\hbeta$ ratio as the $\othree$ lines 
contribute 80~per cent to the $\hbeta$ flux deduced from the P71 filter 
transmission curve. Therefore, we only include their published data  
for completeness, and do not give any weight to the comparison. We notice that, for fluxes measured 
by ASTR91 and KFL88, there tends to be an under-estimation 
of the brightness for PN with decreasing brightness compared to our new fluxes.\\

Note the $\othreec$ fluxes calculated from the recent, more reliable CCD data (R04, JaSt04 and C01) add valuable 
additional data points to test the integrity of the fluxes of fainter PN in the sample. 
Error bars have been included where they exist for both $\halpha$ flux and spectroscopic line ratio. The 
R04 sample has been divided into two groups, one where an error has been calculated from the published flux 
and line ratio and the other where an error has not been calculated due to lack of error information. 
The JaSt04 data is in very close agreement with our data, bar one point with $\Delta\log$F 
$\sim$-0.6. Closer analysis of this individual PN, JaSt65, reveals that this large discrepancy 
is due to it's location in the diffraction halo of a nearby bright star. 
 
The comparison between our fluxes and those derived from $\halpha$ flux data 
is accurate within the errors for PN with fluxes $\lesssim$13.6 in $-\log$F. 
Although it appears errors published in this work for $-\log$F $\gtrsim$13.6 have been under-estimated, 
there is still good agreement with published fluxes up to 15.0 in $-\log$F once 
slightly larger error estimates have been applied. The errors could potentially be as large as 0.2 in 
this, but we adopt 0.09 given the confidence in our data comparison with the best sources at brighter flux limits.\\

Table 3 displays a comparison between our fluxes and those published 
from various literature sources. The horizontal division 
highlights the difference between those published $\othreec$ fluxes that have been measured directly and those 
that used a combination of $\halpha$ or $\hbeta$ flux and spectroscopic line ratios to calculate an 
$\othreec$ flux. The columns list the literature comparison source, the average difference in 
$\Delta\log$F between measurements, the standard deviation of 
errors and how many points are included in the analysis. 
There is a very clear difference between the two comparisons; the 
directly measured photoelectric fluxes and higher quality CCD boot-strapped fluxes can 
clearly be seen to agree far better with our dataset. There is 
a tendency for our fluxes to be slightly brighter than what has been previously published. 
This can be seen most noticably in the comparison of the W83 data, where ours appear constantly offset by 0.03 dex.

\begin{table}
\centering
  \caption{Statistical comparison between fluxes measured here and literature sources. Tabulated are 
the average $\Delta\log$F between measurements, the standard deviation of the differences and how many PN are in the sample. The horizontal division signifies the difference between sources who have published direct 
$\othreec$ fluxes (top), and those where $\othreec$ fluxes have been determined via calculation of the $\halpha$ 
or $\hbeta$ flux and spectroscopic line ratios.}
\begin{tabular}{lllll}
\hline
Reference & $\overline{\Delta\log F}$ & $\sigma_{\Delta\log F}$  & N  & type\\ 
\hline

O63 & -0.01 & 0.02 & 2 & PE \\			
W83 & 0.03 & 0.04 & 16 & PE\\
SK89 & 0.05 & 0.07 & 5 & PE \\
KM81 & 0.01 & - & 1 & PE \\
			
\hline
%Bootstrapped data	
JaSt04 & 0.01 & 0.04 & 7 & CCD \\	 
C01 & -0.09 & 0.09 & 2 & CCD \\
SK89 & 0.05 & 0.20 & 11 & PE\\
P71 & -0.14 & 0.20 & 2 & PE\\
R04 & 0.04 & 0.21 & 14 & CCD \\
KFL88 & 0.11 & 0.21 & 45 & photographic\\
ASTR91 & 0.10 & 0.25 & 71 & CCD\\	
VP85 & 0.19 & - & 1 & photographic\\
\hline
\end{tabular}
   \label{tab:stddevs}
\end{table}

\subsection{Correcting for extinction and provision of dereddened fluxes}
\label{sec:deredden}
During our MASH PN identification
programme follow-up, spectroscopy was obtained for 100~per cent of MASH bulge
PNe. In principle these allow for Balmer decrement measurements and
consequently can provide an estimate of the degree of extinction the $\othreec$ fluxes are subjected to. The AAOmega multi-object
fibre spectroscopy system on the Anglo-Australian telescope (AAT) and
the now decommissioned Double Beam Spectrograph on the MSSSO 2.3m were
primarily used to obtain deep spectra for abundances and 
higher resolution spectra around H$\alpha$ for accurate radial
velocities \citep{M10}. As the multi-object facility on AAOmega allows for observation of 
$\sim$400 objects over a 2-degree field-of-view \citep{L02}, 
fibres were also placed on previously known PN in the 
field enabling observation of most known PN in this region (see Table \ref{tab:fluxes}; see also Fig. 1 of \citealt{M09a}).

Due to their fibre and dual-beam nature
neither type of spectra are ideal for Balmer decrement estimates.
Nevertheless, preliminary dereddened fluxes are presented in Table
\ref{tab:fluxes}. The reader should treat these values with some
caution, and we assign typical errors of $\pm$0.3--0.5~dex at this stage. A
more detailed spectral analysis and improvements to these estimates
are underway as better spectra become available.
Full details of the spectroscopic analysis will be presented in Paper
II, along with the fuller evaluation of the integrity of our
dereddened fluxes.

\section{Angular diameters}
\label{sec:diameters}
As a by-product of this study, angular diameters were calculated for
all observed PN that had detected $\othreec$ emission. Comparison
between PN diameters within a co-located population serves as a proxy for their evolutionary state with those 
that are compact being younger than the more extended PN. Angular diameters can therefore be used to 
investigate whether bulge PN of certain evolutionary states occupy distinct regions in the PNLF.

Major and minor axes were measured using the 10~per cent level of maximum value
method described in detail in \citet{T03}, hereafter T03.  \\

\begin{figure}
\begin{center}
\includegraphics[scale=0.4]{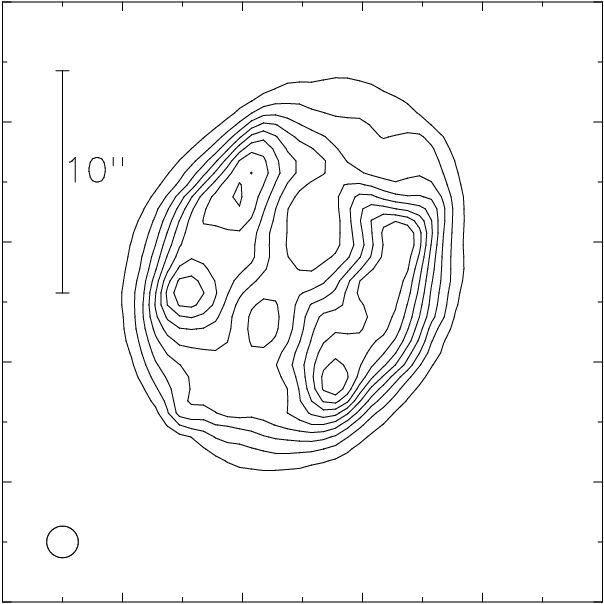}
\includegraphics[scale=0.4]{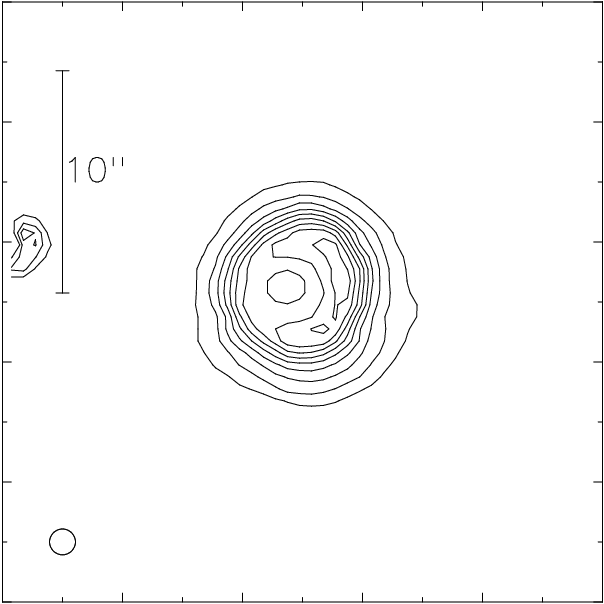}

\caption{$\othreec$ images of PPA1800-2904 (left) and H2-36 (right) with contours at 10~per cent level increments of the
maximum brightness of the PN. The images are 27$\times$27~arcsec, with East up and North to the right. The circles 
in the bottom left hand corners represent the FWHM of the seeing disc. }
\label{fig:contour}
\end{center}
\end{figure}

In practice, the contouring algorithm in ds9 \citep{JM03} 
was used to calculate 10~per cent increments of peak surface brightness, and the
angular diameter was measured at the 10~per cent contour level. Due to the
crowded stellar fields and non-perfect continuum subtraction, it was necessary to measure PN on an individual 
basis to ensure the peak brightness identified was that of the PN and not of an artefact. As the PN 
are often asymmetrical we have measured perpendicular axes, one being along the longest extension of 
emission. Fig. \ref{fig:contour} displays examples of contours calculated for
two PN in our sample: H2-36 and PPA1800-2904. \\

Note that this method is compromised for the compact PN in our sample,
where the angular diameter measurement is comparable to the angular resolution of the instrument, and those few observed in poor seeing.
The observed angular resolution is a combination of the resolution of the camera and the natural seeing.
These contributions needed to be deconvolved to obtain a true measurement of the PN. The full width at half-maximum (FWHM)
of Gaussian fits to stars in each field were used to obtain the FWHM of the instrumental profile. This was then used to deconvolve the instrumental resolution component using equation \eqref{eq:deconv} given in T03:

\begin{equation}
\label{eq:deconv}
\Theta_{10~per cent,d} = \sqrt{\Theta^{2}_{10~per cent} - (1.823\Phi_{b})^2}
\end{equation}

where $\Phi_{b}$ is the FWHM of the resolution profile of the instrument, $\Theta_{10~per cent}$ is the angular diameter of the PN at the 10~per cent level and $\Theta_{10~per cent,d}$ is the deconvolved PN diameter at the 10~per cent level. \\

\subsection{Comparison of duplicate observations of PN}

The errors in the measured angular diameters of PN were determined via comparison between 
duplicate major axes measurements for PN observed in more than one epoch during our survey. The difference in repeat measurements 
of a PN angular diameter ($\Delta$ D), is seen to be independent of measured diameter, 
so the standard deviation of the distribution ($0.75$~arcsec) is applied to all measurements as an estimate of the error.

\begin{figure}
\begin{center}
\includegraphics[scale=0.6]{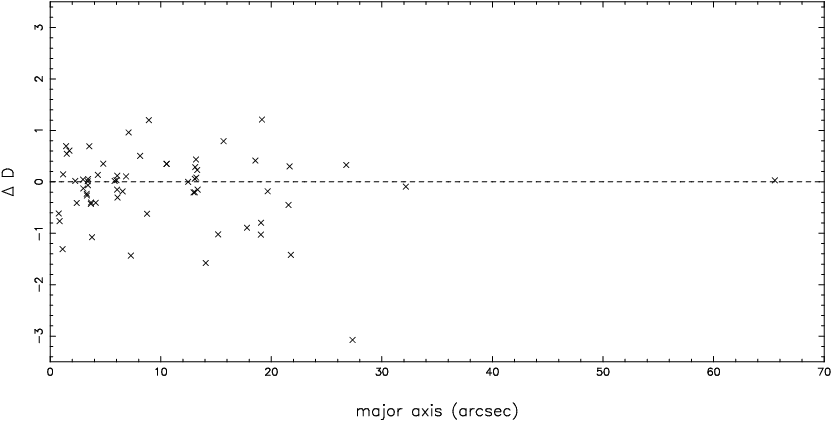}

\caption{A comparison of angular diameters for 62 PN with duplicate measurements. The difference between duplicate 
measurements is seen to be independent of the major axis measured. 
The standard deviation of the distribution is the error assigned to all angular diameter measurements. }
\label{fig:errDiam}
\end{center}
\end{figure}

\subsection{Comparison with literature diameters}
\begin{figure}
\centering
\includegraphics[scale=0.6]{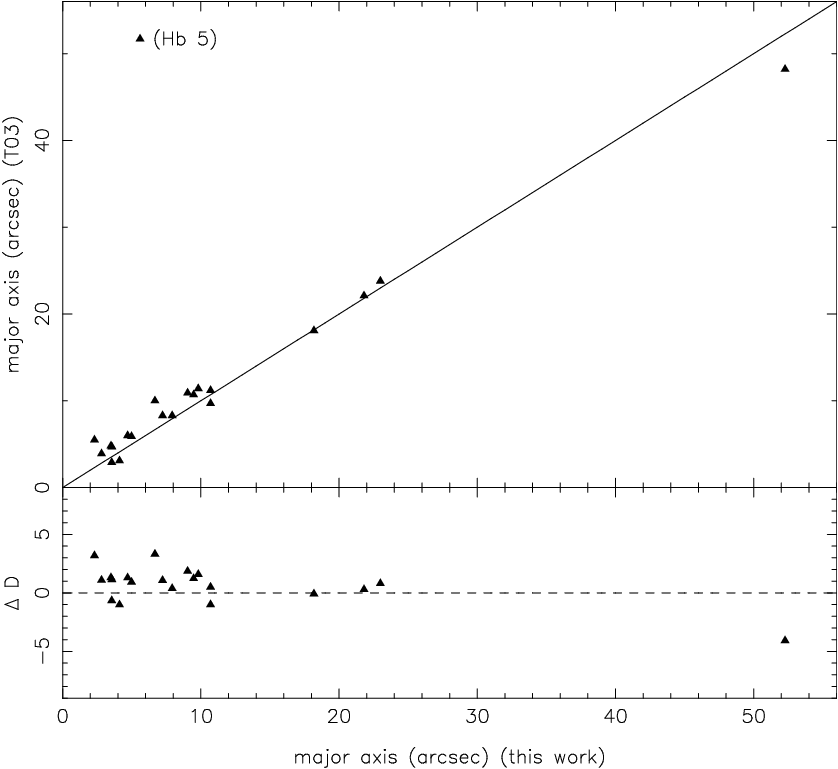}

\caption{Our measured angular diameters compared against those from 
Tylenda et al. (2003). There is very good agreement exept for data-point in the top left
hand corner. This object is Hb 5, whose $\othreec$ 10~per cent contour level delineates 
the bright core as opposed to the $\halpha$ diameter which 
encompasses the outer bipolar structure.}
\label{fig:diamComp}
%\end{center}
\end{figure}

Prior to this work one of the most recent and homogenous compilations of bulge PN 
angular diameters was by T03 though their measurements are from $\halpha$ images. 
However, provided the PNe can be considered optically
thick, then the $\halpha$ and $\othreec$ ionisation regions are comparable, so a direct comparison 
between their $\halpha$ and our $\othreec$ angular diameters can be made as shown in Fig. \ref{fig:diamComp}. 
The majority of points are in good agreement apart from that of Hb~5. Hb~5 is a strongly bipolar 
nebula that is of similar morphology to the Homunculus around Eta Carinae. T03 note that 
such large discrepancies are common in 
extended nebulae with bright bipolar cores. We therefore urge the reader to 
exercise caution when dealing with similar types of PN. In the case of Hb~5, 
their $\halpha$ measurement encompasses a significant portion of outer bipolar structures,
whereas our $\othreec$ diameter measurement only delineates the bright core. 
The smaller diameter in $\othreec$ is a consequence of ionisation stratification. \\

\section{Conclusions}
In this paper, we have increased the number of PN with directly measured $\othreec$ 
fluxes in the $10\arcdeg\times10\arcdeg$ toward the Galactic bulge region by a factor of eight, 
with the addition of 387 previously unobserved PN. We have also re-observed 48 PN with directly 
measured $\othreec$ fluxes from the literature. 
Our new fluxes account for $\sim$80\% of known PN towards 
the Galactic bulge. We have provided additional fluxes for five PN in 
the peripheral regions, and for a special PN outside the bulge region (PHR1315-6555) 
that is the subject of another paper \citep{P10}.
This brings the total number of PN for which we have provided 
measured fluxes for to 441. \\

Photometric errors were derived via consideration of the S/N measurement, the velocities of the PN, 
the transmission profile of the filter curve and those due to the variation in the sensitivity 
factor calculated from the standard stars used. We found that these errors under-estimated the true 
error after comparison between duplicate measurements 
of a PN, so we increased the errors accordingly. 
We have compared our fluxes to all 
previously published $\othreec$, $\othreeb$, $\halpha$ and $\hbeta$ fluxes, 
and found there 
to be no statistically significant systematic offset between the most trusted 
datasets and our new flux measurements to within the errors. Angular diameter measurements for 
all PN observed are also included, along with preliminary estimates of the dereddened fluxes based on the current best available spectra. \\

These data therefore provide accurate fluxes and angular diameters for the largest sample of PN 
in the $10\arcdeg\times10\arcdeg$ region toward the Galactic bulge ever collated. 
Deriving from the same high quality, uniform, 
high resolution \linebreak MOSAIC-II imaging, precise angular diameter measurements and greater insight 
into the fine morphological detail of each PN can be made. Combination of these factors will allow for 
the most detailed and accurate construction of the bulge PNLF to date. This will be the subject 
of Paper II.

\section*{Acknowledgments}
AVK acknowledges Macquarie University for a PhD scholarship. 
QAP and AVK wish to thank ANSTO for the travel grants which facilitated both 
observing trips to the CTIO BLANCO 4-m to collect this data. We thank NOAO 
for the award of telescope time. We 
acknowledge and thank Griet Van de Steene for allowing access to unpublished 
spectroscopic data for the JaSt PN. This paper has made use 
of the Simbad and Vizier database services, as well as the Aladin 
software tools. We thank Thomas Boch for writing the Aladin instrument FOV template for 
the MOSAIC-II camera. Thanks to Francois Ochsenbein for help with
the PN G numbers. We thank an anonymous referee for some very valuable and useful 
comments.

\appendix \section{Field centres and Fluxes}
\clearpage
%\begin{landscape}
\begin{table}
\caption{A list of the date field centres that were observed. All co-ordinates are J2000. }
\begin{tabular}{lllll}
\hline
Field name & RA & Dec & seeing & exposure time \\
(1) & (2) & (3) & (4) & (5) \\
\hline
\multicolumn{5}{|c|}{9/6/2008} \\
F1753$-$2841 & 17 53 52.1 & $-$28 41 15 &2.04 & 3x400s \\
F1754$-$2915 & 17 54 04.8 & $-$29 15 15 &1.39 & 2x400s \\ 
F1751$-$2949 & 17 51 41.3 & $-$29 49 15 &1.26& 3x400s \\ 
F1754$-$2949 & 17 54 17.9 & $-$29 49 15 &1.32&  1x60s;3x400s \\ 
F1758$-$2733 & 17 58 34.3 & $-$27 33 15 &1.29& 3x400s \\ 
F1804$-$2841 & 18 04 11.8 & $-$28 41 15 &1.26 & 1x60s;3x400s \\ 
F1759$-$2915 & 17 59 16.4 & $-$29 15 15 &1.16& 3x400s \\ 
F1801$-$2841 & 18 01 36.9 & $-$28 41 15 &1.23& 1x60s;3x400s \\ 
F1811$-$2807 & 18 11 38.5 & $-$28 07 15 &1.26& 1x60s;3x400s \\ 
F1801$-$2807 & 18 01 22.0 & $-$28 07 15 &1.21& 1x60s;3x400s \\ 
F1759$-$2841 & 17 59 01.9 & $-$28 41 15 &1.30& 3x400s \\ 
F1802$-$3131 & 18 02 57.8 & $-$31 31 15 &1.42& 1x60s;3x400s \\ 
F1803$-$2807 & 18 03 56.1 & $-$28 07 15 &1.52& 3x400s \\
\hline
\multicolumn{5}{|c|}{10/6/2008} \\
F1719$-$3048 & 17 19 44.8 & $-$30 48 39 &1.55&  3x400s \\ 
F1751$-$2841 & 17 51 17.1 & $-$28 41 15 &1.33& 3x400s \\ 
F1751$-$2915 & 17 51 29.0 & $-$29 15 15 &1.28& 3x400s \\ 
F1753$-$2807 & 17 53 39.7 & $-$28 07 15 &1.42& 3x400s \\ 
F1756$-$2841 & 17 56 27.0 & $-$28 41 15 &1.63 &3x400s \\ 
F1756$-$2915 & 17 56 40.6 & $-$29 15 15 &1.27 &3x400s \\ 
F1751$-$3023 & 17 51 53.8 & $-$30 23 15 &1.33 &3x400s \\ 
F1806$-$2807 & 18 06 30.3 & $-$28 07 15 &1.23 &3x400s \\ 
F1756$-$2807 & 17 56 13.8 & $-$28 07 15 &1.36 &3x400s \\ 
F1758$-$2807 & 17 58 47.9 & $-$28 07 15 &1.29 &1x60s;3x400s \\ 
F1756$-$2733 & 17 56 01.0 & $-$27 33 15 &1.25 &3x400s \\ 
F1801$-$2733 & 18 01 07.6 & $-$27 33 15 &1.31 &3x400s \\ 
F1806$-$2715 & 18 06 54.4 & $-$27 15 54 &1.45 &3x400s \\ 
F1811$-$2722 & 18 11 18.8 & $-$27 22 35 &1.45 &1x60s;3x400s \\ 
\hline
\multicolumn{4}{|c|}{11/6/2008} \\
F1725$-$3055 & 17 25 26.8 & $-$30 55 00 &1.22 &3x400s \\ 
F1804$-$2703 & 18 04 22.8 & $-$27 03 33 &1.14 &1x60s;3x400s \\ 
F1806$-$2642 & 18 06 56.6 & $-$26 42 03 &1.15 &1x30s;3x400s \\ 
F1801$-$2915 & 18 01 52.1 & $-$29 15 15 &1.24 &3x400s \\ 
F1804$-$2915 & 18 04 27.9 & $-$29 15 15 &1.24 &3x400s \\ 
F1807$-$2907 & 18 07 03.7 & $-$29 07 21 &1.12 &1x30s;3x400s \\ 
F1807$-$2833 & 18 07 19.5 & $-$28 33 20 &1.09 &1x30s;3x400s \\ 
F1752$-$3123 & 17 52 20.1 & $-$31 23 21 &1.10 &3x400s \\ 
F1749$-$3015 & 17 49 16.3 & $-$30 15 21 &1.02 &3x400s \\ 
F1749$-$3049 & 17 49 28.3 & $-$30 49 21 &1.23 &1x30s;3x400s \\ 
F1804$-$2841 & 18 04 11.8 & $-$28 41 15 &1.21 &1x60s;2x400s \\ 
F1806$-$2715 & 18 06 54.4 & $-$27 15 54 &1.22 &2x400s \\ 
\hline
\multicolumn{4}{|c|}{12/6/2008} \\
F1718$-$3006 & 17 18 37.7 & $-$30 06 30 &1.19 &3x400s \\ 
F1800$-$2651 & 18 00 04.2 & $-$26 51 17 &1.22 &3x400s \\ 
F1755$-$3025 & 17 55 04.8 & $-$30 25 01 &1.20 &1x30s;3x400s \\ 
F1757$-$3115 & 17 57 34.9 & $-$31 15 14 &1.16 &3x400s \\ 
F1754$-$3115 & 17 54 56.2 & $-$31 15 02 &1.14 &3x400s \\ 
F1757$-$3149 & 17 57 12.9 & $-$31 49 36 &0.96 &3x400s \\ 
F1758$-$3314 & 17 58 59.4 & $-$33 14 21 &0.96 &1x30s;3x400s \\ 
F1751$-$3327 & 17 51 15.6 & $-$33 27 01 &0.99 &3x400s \\ 
F1752$-$3401 & 17 52 58.4 & $-$34 01 00 &1.00 &3x400s \\ 
F1750$-$3400 & 17 50 14.2 & $-$34 00 54 &1.03 &3x400s \\ 
F1753$-$3435 & 17 53 52.8 & $-$34 35 10 &1.08 &1x30s;3x400s \\ 
F1753$-$3518 & 17 53 24.9 & $-$35 18 01 &1.16 &3x400s \\ 
\hline
\multicolumn{4}{|c|}{13/6/2008} \\ 
F1722$-$2855 & 17 22 17.8 & $-$28 55 39 &1.63 &1x30s;3x400s \\ 
F1740$-$3318 & 17 40 27.0 & $-$33 18 58 &1.48 &3x400s \\ 
F1744$-$3302 & 17 44 06.9 & $-$33 02 54 &1.09 &3x400s \\ 
\hline
\end{tabular}
\label{tab:fields}
\end{table}
%\end{landscape}

%\begin{landscape}
\begin{table}
\contcaption{} 
\begin{tabular}{lllll}\\
\hline
Field name & RA & Dec &seeing & exposure time \\
(1) & (2) & (3) & (4) & (5) \\
\hline
\multicolumn{5}{|c|}{13/6/2008} \\ 
F1743$-$3400 & 17 43 10.3 & $-$34 00 44 &1.28 &1x30s;3x400s \\ 
F1752$-$3243 & 17 52 50.4 & $-$32 43 53 &1.19 &1x30s;3x400s \\ 
F1743$-$2725 & 17 43 14.5 & $-$27 25 21 &1.05 &3x400s \\ 
F1745$-$3428 & 17 45 36.6 & $-$34 28 18 &1.02 &1x30s;3x400s \\ 
F1750$-$3203 & 17 50 43.4 & $-$32 03 53 &1.23 &3x400s \\ 
F1800$-$2958 & 18 00 24.5 & $-$29 58 21 &1.05 &3x400s \\ 
F1802$-$3049 & 18 02 52.8 & $-$30 49 04 &1.20 &3x400s \\ 
F1803$-$2940 & 18 03 02.1 & $-$29 40 31 &1.10 &1x30s;3x400s \\ 
F1805$-$3048 & 18 05 54.8 & $-$30 48 24 &1.19 &1x30s;3x400s \\ 
F1800$-$3032 & 18 00 22.1 & $-$30 32 36 &1.15 &3x400s \\ 
F1805$-$3014 & 18 05 38.4 & $-$30 14 17 &1.25 &3x400s \\ 
F1811$-$2807 & 18 11 38.5 & $-$28 07 15 &1.40 &1x60s;2x400s \\ 
F1811$-$2722 & 18 11 18.8 & $-$27 22 35 &1.48 &1x60s;2x400s \\ 
\hline
\multicolumn{5}{|c|}{14/6/2008} \\
F1722$-$3209 & 17 22 18.0 & $-$30 29 41 &1.81 &3x400s \\ 
F1746$-$3200 & 17 46 58.8 & $-$32 00 34 &1.58 &3x400s \\ 
F1802$-$3223 & 18 02 21.0 & $-$32 23 31 &1.29 &3x400s \\ 
F1746$-$3325 & 17 46 47.6 & $-$33 25 08 &1.19 &1x30s;3x400s \\ 
F1748$-$3450 & 17 48 04.4 & $-$34 50 45 &1.19 &3x400s \\ 
F1751$-$3444 & 17 51 07.5 & $-$34 44 17 &1.18 &1x30s;3x400s \\ 
F1757$-$3014 & 17 57 43.8 & $-$30 14 17 &1.60 &1x30s;3x400s \\ 
F1800$-$2511 & 18 00 10.6 & $-$25 11 37 &1.42 &1x30s;3x400s \\ 
F1803$-$3014 & 18 03 01.5 & $-$30 14 37 &1.27 &1x30s;3x400s \\ 
F1757$-$2940 & 17 57 08.4 & $-$29 40 26 &1.32 &1x30s;3x400s \\ 
F1759$-$3210 & 17 59 47.2 & $-$32 10 04 &0.94 &1x30s;3x400s \\ 
F1752$-$3049 & 17 52 06.8 & $-$30 49 21 &1.14 &1x30s;3x400s \\ 
F1746$-$3049 & 17 46 49.8 & $-$30 49 21 &1.15 &3x400s \\ 
F1800$-$3106 & 18 00 14.7 & $-$31 06 08 &1.10 &3x400s \\ 
F1811$-$2833 & 18 11 23.9 & $-$28 33 37 &1.26 &1x30s;3x400s \\
\hline
\multicolumn{5}{|c|}{27/6/2009} \\
F1748$-$2429 & 17 48 42.3 & $-$24 29 29 &2.68 &3x400s \\ 
F1737$-$2802 & 17 37 51.7 & $-$28 02 20 &3.09 &1x30s;3x400s \\ 
F1740$-$2759 & 17 40 29.5 & $-$27 59 57 &3.23 &3x400s \\ 
F1730$-$3009 & 17 30 43.6 & $-$30 09 06 &2.00 &3x400s \\ 
F1742$-$2542 & 17 42 40.9 & $-$25 42 52 &2.48 &3x400s \\ 
F1736$-$2910 & 17 36 00.3 & $-$29 10 32 &2.69 &3x400s \\ 
F1725$-$2929 & 17 25 18.5 & $-$29 29 35 &2.99 &1x30s;3x400s \\ 
\hline
\multicolumn{5}{|c|}{28/6/2009} \\ 
F1809$-$2905 & 18 09 58.7 & $-$29 05 06 &2.69 &1x30s;3x400s \\ 
F1804$-$2637 & 18 04 16.7 & $-$26 37 37 &2.68 &1x30s;3x400s \\ 
\hline
\multicolumn{5}{|c|}{29/6/2009} \\ 
F1745$-$2619 & 17 45 38.3 & $-$26 19 09 &2.24 &3x400s \\ 
F1735$-$2754 & 17 35 25.2 & $-$27 54 27 &2.15 &3x400s \\ 
F1745$-$2543 & 17 45 03.6 & $-$25 43 57 &1.71 &1x30s;3x400s \\ 
F1743$-$2651 & 17 43 03.6 & $-$26 51 06 &1.73 &5x400s \\ 
F1739$-$2834 & 17 39 55.3 & $-$28 34 54 &1.74 &3x400s \\ 
F1733$-$2942 & 17 33 28.9 & $-$29 42 00 &1.58 &1x30s;3x400s \\ 
\hline
\multicolumn{5}{|c|}{1/7/2009} \\ 
F1728$-$2755 & 17 28 49.5 & $-$27 55 39 &1.96 &3x400s \\ 
F1731$-$2821 & 17 31 18.7 & $-$28 21 02 &1.89 &1x30s;3x400s \\ 
F1730$-$2857 & 17 30 43.4 & $-$28 57 54 &2.04 &1x30s;3x400s \\ 
F1728$-$2833 & 17 28 34.5 & $-$28 33 30 &1.81 &1x30s;3x400s \\ 
F1740$-$2650 & 17 40 22.2 & $-$26 50 05 &2.02 &2x400s \\ 
F1734$-$2641 & 17 34 09.7 & $-$26 41 15 &1.83 &3x400s \\ 
F1729$-$2603 & 17 29 37.3 & $-$26 03 39 & 2.08&3x400s \\ 
F1735$-$3019 & 17 35 30.6 & $-$30 19 02 &1.96 &2x400s \\ 
F1737$-$2835 & 17 37 32.7 & $-$28 35 38 & 1.85&2x400s \\  
\hline
\end{tabular}
\end{table}
%\end{landscape}

%\begin{landscape}
\begin{table}
\contcaption{} 
\begin{tabular}{|lllll|}
\hline
Field name & RA & Dec &seeing & exposure time \\
(1) & (2) & (3) & (4) & (5)\\
\hline
\multicolumn{5}{|c|}{1/7/2009} \\ 
F1732$-$3015 & 17 32 58.5 & $-$30 15 18 &1.81 &2x400s \\
F1743$-$2651 & 17 43 03.6 & $-$26 51 06 & 1.76& 1x100s \\ 
F1745$-$3401 & 17 45 23.3 & $-$34 01 05 & 1.68&3x400s \\ 
F1810$-$2759 & 18 10 21.0 & $-$27 59 08 & 1.62&3x400s \\ 
\multicolumn{5}{|c|}{2/7/2009} \\
F1736$-$2328 & 17 36 17.8 & $-$23 28 58 & 1.81&2x400s \\ 
F1740$-$2426 & 17 40 54.1 & $-$24 26 53 & 1.73&1x30s;3x400s \\ 
F1738$-$2911 & 17 38 34.3 & $-$29 11 29 & 1.71&3x400s \\ 
F1735$-$2945 & 17 35 57.8 & $-$29 45 19 & 1.70&3x400s \\ 
F1740$-$2723 & 17 40 32.8 & $-$27 23 56 & 1.53&3x400s \\ 
F1753$-$3435 & 17 53 52.8 & $-$34 35 10 & 1.23&1x100s \\ 
F1748$-$3415 & 17 48 05.4 & $-$34 15 19 & 1.20&1x30s;3x400s \\  
F1756$-$3424 & 17 56 22.4 & $-$34 25 58 & 1.26&1x30s;3x400s \\ 
F1757$-$3349 & 17 57 39.8 & $-$33 49 22 & 1.31&1x30s;3x400s \\ 
F1743$-$2419 & 17 43 31.1 & $-$24 19 33 & 1.37&1x30s;3x400s \\ 
F1748$-$2344 & 17 48 40.0 & $-$23 44 33 & 1.39&2x400s \\ 
F1730$-$2722 & 17 30 58.4 & $-$27 22 14 & 1.36&3x400s \\ 
F1737$-$2513 & 17 37 42.7 & $-$25 13 56 & 1.53&3x400s \\ 
F1735$-$2711 & 17 35 27.1 & $-$27 11 48 & 1.52&1x400s \\ 
F1810$-$2759 & 18 10 21.0 & $-$27 59 08 & 1.78&1x30s \\ 
F1808$-$2938 & 18 08 07.6 & $-$29 38 56 & 1.58&1x400s \\
\hline
\end{tabular}
\end{table}

\begin{landscape}
\begin{table}
\caption{A table depicting our $\othreec$ global fluxes for 435 PNe in the $10\arcdeg\times10\arcdeg$
region toward the Galactic Bulge. The PN G name 
and their common name are listed in the first two columns, the observed $\othreec$ flux 
measured directly from the CCD image is presented in column (3) (in erg cm$^{-2}$s$^{-1}$) with their 
estimated errors (4), with their preliminary dereddened fluxes in (5), and their major and minor axes are listed in columns (6) and (7). 
Diameters are subject to a $0.75\arcsec$ error, and dereddened fluxes are subject to an error 
of $\pm$0.3--0.5~dex.
%All fluxes presented are uncorrected for interstellar absorption. 
}
\label{tab:fluxes}
\begin{minipage}{4.5in}
 \begin{tabular}{|l l l l l l l|}
\hline
 & & \multicolumn{3}{|c|}{$-\log$(F[OIII])} &  \multicolumn{2}{|c|}{Diameter ($\arcsec$)}\\ 
PN G name & Usual name & Flux & error & F$_{dered}^d$ & maj & min \\
(1) & (2) & (3) & (4) & (5) & (6) & (7) \\
\hline
000.0$-$01.3 & PPA1751-2933 & ND$^b$ & -- & -- & -- & -- \\ 
000.0$-$01.8 & JaSt83 & 13.00 & 0.03 & 10.77 & 3.3 & 3.3 \\ 
000.0$-$01.8a & PHR1752-2953 & 13.26 & 0.03 & -- & 17.8 & 16.8 \\ 
000.0$-$02.1 & MPA1754-2957 & 13.05 & 0.03 & 11.09 & 21.5 & 19.0 \\ 
000.0$-$02.5 & K6-36 & 12.91 & 0.03 & 11.06 & 10.5 & 9.8 \\ 
000.0$+$01.3 & JaSt27 & ND$^b$ & -- & 12.85 & -- & -- \\ 
000.0$+$02.0 & K6-4 & 14.97 & 0.21 & 10.96 & 6.9 & 4.1 \\ 
000.0$-$02.9 & MPA1757-3021 & ND$^a$$^b$ & -- & -- & -- & -- \\ 
000.1$-$01.0 & JaSt69 & 13.90 & 0.09 & 10.65 & 11.1 & 9.4 \\ 
000.1$-$01.1 & M3-43 & 12.88 & 0.03 & 9.06 & 3.5 & 2.7 \\ 
000.1$-$01.2 & JaSt75 & 14.37 & 0.09 & 9.47 & 2.9 & 1.9 \\ 
000.1$-$01.7 & PHR1752-2941 & 12.52 & 0.03 & 10.31 & 16.7 & 12.2 \\ 
000.1$-$01.9 & JaSt93 & 12.41 & 0.02 & 10.37 & 14.2 & 11.7 \\ 
000.1$+$01.9 & PHR1738-2748 & 13.90 & 0.09 & 10.77 & 13.6 & 9.8 \\ 
000.1$+$02.0 & PHR1738-2746 & 13.27 & 0.03 & -- & 18.9 & 17.5 \\ 
000.1$+$02.6 & Al2-J & 12.61 & 0.03 & 10.46 & 11.4 & 8.9 \\ 
000.2$-$01.9 & M2-19 & 12.22 & 0.02 & 10.53 & 6.7 & 6.1 \\ 
000.2$-$01.9a & JaSt2-14 & ND$^b$ & -- & -- & -- & -- \\ 
000.2$-$02.3 & Bl3-10 & 11.64 & 0.02 & 10.13 & 7.2 & 6.9 \\ 
000.2$-$03.4 & PHR1759-3030 & 12.91 & 0.03 & 11.95 & 21.2 & 18.2 \\ 
000.2$-$04.6 & Sa3-117 & 12.51 & 0.03 & 11.35 & 8.2 & 6.8 \\ 
000.2$+$01.4 & PBOZ26 & 14.14 & 0.09 & -- & 3.2 & 3.0 \\ 
000.2$+$01.7 & JaSt19 & 13.09 & 0.03 & 10.06 & 7.2 & 6.4 \\
000.2$-$04.0 & MPA1802-3045 & ND$^b$ & -- & -- & -- & -- \\ 
000.3$-$01.6 & PHR1752-2930 & 12.93 & 0.03 & 10.13 & 8.6 & 7.9 \\ 
000.3$-$02.8 & M3-47 & 12.60 & 0.03 & 11.24 & 8.1 & 7.2 \\ 
000.3$+$01.5 & JaSt23 & ND$^b$ & -- & -- & -- & -- \\ 
000.3$+$01.7 & JaSt21 & 14.00 & 0.09 & -- & 19.2 & 18.9 \\ 
000.3$+$03.2 & PHR1733-2655 & 13.09 & 0.03 & -- & 20.8 & 10.9 \\ 
000.3$+$03.4 & PHR1733-2647 & 13.58 & 0.09 & -- & 16.8 & 15.0 \\ 
000.3$+$04.2 & PPA1730-2621 & 13.79 & 0.09 & -- & 7.6 & 7.3 \\ 
000.3$+$04.5 & PHR1729-2614 & 13.66 & 0.09 & -- & 14.8 & 10.5 \\ 
000.3$-$03.4 & MPA1800-3023 & 13.24 & 0.03 & 12.13 & 7.9 & 5.8 \\ 
000.3$-$04.2 & MPA1803-3046 & 13.74 & 0.09 & 12.94 & 23.7 & 20.5 \\ 
000.3$-$04.2a & MPA1803-3043 & 13.32 & 0.03 & 12.52 & 36.6 & 36.2 \\ 
000.4$-$01.3 & JaSt2-8 & ND$^b$ & -- & -- & --  \\ 
000.4$-$01.9 & M2-20 & 11.54 & 0.02 & 10.15 & 3.6 & 3.2 \\ 
000.4$-$02.9 & M3-19 & 11.90 & 0.02 & 10.59 & 7.2 & 6.6 \\ 
000.4$+$04.4 & PPA1729-2611 & 12.85 & 0.03 & -- & 6.4 & 5.5 \\ 
000.5$-$01.5 & JaSt2-9 & 14.14 & 0.09 & 10.76 & 8.9 & 6.1 \\ 
000.5$-$01.6 & Al2-Q & 12.56 & 0.03 & 10.39 & 7.8 & 7.6 \\ 
000.5$-$01.7 & JaSt96 & 12.98 & 0.03 & 11.46 & 32.9 & 31.2 \\
000.5$-$03.1 & KFL1 & 12.14 & 0.02 & 10.64 & 8.0 & 7.9 \\ 
\hline
    \end{tabular}
  \end{minipage}
\begin{minipage}{4.5in}  
\begin{tabular}{|l l lllll|}
\hline
 & & \multicolumn{3}{|c|}{$-\log$(F[OIII])} &  \multicolumn{2}{|c|}{Diameter ($\arcsec$)}\\ 
PN G name & Usual name & Flux & error & F$_{dered}^d$ & maj & min \\
(1) & (2) & (3) & (4) & (5) & (6) & (7) \\
\hline 
000.5$+$01.9 & JaSt17 & 13.23 & 0.03 & 9.89 & 9.1 & 6.6 \\ 
000.5$+$02.8 & PHR1735-2659 & 12.76 & 0.03 & -- & 26.5 & 25.6 \\ 
000.5$-$03.1 & MPA1759-3007 & 14.45 & 0.10 & 13.48 & 17.9 & 8.1 \\ 
000.6$-$01.0 & JaSt77 & 13.43 & 0.03 & 7.89 & 10.3 & 9.6 \\ 
000.6$-$01.3 & Bl3-15 & 15.39 & 0.22 & 12.47 & 3.9 & 1.2 \\ 
000.6$-$01.4 & PHR1753-2905 & 13.37 & 0.03 & 11.30 & 14.5 & 14.5 \\ 
000.6$-$01.6 & MPA1753-2916 & ND$^a$$^b$ & -- & -- & -- & -- \\ 
000.6$-$04.5 & PM1-206 & 13.04 & 0.03 & 11.50 & 20.3 & 17.3 \\ 
000.6$+$03.1 & PHR1734-2644 & 13.36 & 0.03 & -- & 16.7 & 15.5 \\ 
000.6$-$03.2 & MPA1759-3004 & 12.59 & 0.03 & -- & 30.2 & 30.1 \\ 
000.7$-$00.8 & JaSt74 & ND$^b$ & -- & -- & -- & -- \\ 
000.7$-$01.5 & JaSt2-11 & 13.80 & 0.09 & 11.47 & 9.7 & 8.8 \\ 
000.7$-$02.7 & M2-21 & 10.96 & 0.02 & -- & 2.8 & 2.8 \\ 
000.7$-$03.7 & M3-22 & 11.63 & 0.02 & 10.81 & 11.7 & 7.9 \\ 
000.7$+$03.2 & He2-250 & 12.29 & 0.02 & -- & 6.7 & 4.9 \\ 
000.7$+$04.7 & H2-11 & 12.99 & 0.03 & -- & 3.6 & 3.4 \\ 
000.8$-$00.6 & JaSt71 & ND$^b$ & -- & -- & -- & -- \\ 
000.8$-$01.5 & BlO & ND$^b$ & -- & -- & -- & -- \\ 
000.8$+$01.3 & JaSt38 & 14.36 & 0.09 & -- & 10.9 & 9.6 \\ 
000.9$-$01.2 & JaSt84 & 14.06 & 0.09 & 11.72 & 13.8 & 3.6 \\ 
000.9$-$01.4 & JaSt2-13 & 14.12 & 0.10 & -- & 26.5 & 25.9 \\ 
000.9$-$02.0 & Bl3-13 & 12.02 & 0.02 & 10.27 & 4.2 & 3.9 \\ 
000.9$-$03.3 & PHR1801-2947 & 12.97 & 0.03 & 11.72 & 35.1 & 31.2 \\ 
000.9$-$04.2 & PHR1804-3016 & 12.61 & 0.03 & 11.82 & 11.5 & 11.3 \\ 
000.9$-$04.8 & M3-23 & 11.09 & 0.02 & -- & 13.6 & 12.5 \\ 
000.9$+$01.1 & JaSt44 & 14.24 & 0.09 & -- & 8.5 & 5.0 \\ 
000.9$+$01.8 & PPA1740-2708 & 14.10 & 0.09 & 10.67 & 3.4 & 3.1 \\ 
000.9$+$02.1 & MPA1739-2702 & 13.60 & 0.09 & -- & 19.0 & 11.9 \\ 
001.0$-$01.2 & JaSt87 & 13.92 & 0.09 & -- & 25.4 & 17.6 \\ 
001.0$-$01.5 & MPA1754-2847 & 14.47$^a$ & 0.09 & -- & 15.7 & 11.1 \\ 
001.0$-$01.9 & PHR1755-2904 & 11.90 & 0.02 & 10.26 & 20.3 & 14.6 \\ 
001.0$-$02.6 & Sa3-104 & 12.84 & 0.03 & 11.00 & 1.7 & 1.3 \\ 
001.0$+$01.3 & JaSt41 & 13.15 & 0.03 & 10.88 & 4.7 & 4.6 \\ 
001.0$+$01.4 & JaSt2-4 & ND$^b$ & -- & -- & -- & -- \\ 
001.0$+$01.9 & K1-4 & 11.51 & 0.02 & -- & 52.3 & 47.4 \\ 
001.0$+$02.2 & PPA1739-2652 & ND$^b$ & -- & -- & -- & -- \\ 
001.0$+$02.2 & MPA1739-2652 & 13.66$^a$ & 0.09 & -- & 21.9 & 21.4 \\ 
\hline
    \end{tabular}
\ntd{$^a$This paper includes fluxes for several additional new MASH-II PN to be published in \citet{M10}.}
\ntd{$^b$ND = No $\othreec$ flux was detected.}
\ntd{$^c$HS = Seeing was too high to allow for accurate diameter measurement.}
\ntd{$^d$ Preliminary estimate of dereddened $\othreec$ flux with typical errors of $\pm 0.2$~dex based on the current best available spectra.}
  \end{minipage}
 \end{table}
\end{landscape}
%------------------------------------------------------------------------------------------------
\begin{landscape}
\begin{table}
\contcaption{ }
\begin{minipage}{4.5in}
 \begin{tabular}{lllllll}
\hline
 & & \multicolumn{3}{|c|}{$-\log$(F[OIII])} &  \multicolumn{2}{|c|}{Diameter ($\arcsec$)}\\ 
PN G name & Usual name & Flux & error & F$_{dered}^d$ & maj & min \\
(1) & (2) & (3) & (4) & (5) & (6) & (7) \\
\hline
001.1$-$01.2 & PPA1753-2836 & 12.83 & 0.03 & 10.05 & 10.3 & 8.5 \\ 
001.1$-$01.6 & Al2-S & 12.59 & 0.03 & 10.39 & 6.4 & 5.7 \\ 
001.1$-$02.6 & MPA1758-2915 & 12.95$^a$ & 0.03 & -- & 24.4 & 22.9 \\ 
001.1$+$01.6 & JaSt34 & ND$^b$ & -- & -- & --  \\ 
001.1$+$02.2 & MPA1739-2648 & 14.53 & 0.09 & 10.90 & 4.5 & 3.8 \\ 
001.2$-$01.2 & JaSt92 & 14.22 & 0.09 & -- & 9.5 & 4.7 \\ 
001.2$-$01.2a & JaSt95 & 13.05 & 0.03 & 9.94 & 10.3 & 8.6 \\ 
001.2$-$01.4 & JaSt2-15 & 14.05 & 0.11 & -- & -- & -- \\ 
001.2$-$02.0 & PHR1756-2857 & 13.90 & 0.09 & -- & 22.7 & 17.0 \\ 
001.2$-$02.6 & PHR1759-2915 & ND$^b$ & --& --  & -- & -- \\ 
001.2$-$03.0 & H1-47 & 14.18 & 0.09 & 12.55 & 1.1 & 0.9 \\ 
001.2$-$03.8 & PHR1803-2947 & 12.70 & 0.03 & -- & 48.7 & 48.1 \\ 
001.2$-$03.9 & KFL5 & 11.76 & 0.02 & 10.73 & 1.3 & 1.1 \\ 
001.2$+$01.3 & JaSt45 & 13.98 & 0.09 & -- & 31.3 & 30.6 \\ 
001.2$+$02.1 & He2-262 & 12.35 & 0.02 & 10.07 & 4.6 & 4.5 \\ 
001.3$-$01.0 & JaSt89 & 13.57 & 0.09 & -- & 9.5 & 6.7 \\ 
001.3$-$01.2 & Bl1-M & 14.08 & 0.09 & 10.88 & 3.5 & 3.5 \\ 
001.4$-$04.4 & BMP1806-2958 & 12.89 & 0.03 & --  & 37.6 & 35.6 \\ 
001.4$-$03.4 & ShWi1 & 12.45 & 0.02 & 9.63 & 16.3 & 15.2 \\ 
001.5$-$01.1 & JaSt2-12 & ND$^b$ & -- & -- & -- & -- \\ 
001.5$-$01.6 & PHR1755-2825 & 14.52 & 0.09 & -- & 13.0 & 1.2 \\ 
001.5$-$01.8 & JaSt2-19 & 14.20 & 0.09 & 11.14 & 5.3 & 2.7 \\ 
001.5$-$02.4 & PHR1758-2852 & 12.70 & 0.03 & 11.23 & 17.2 & 14.4 \\ 
001.5$-$02.8 & PPA1800-2904 & 12.45 & 0.02 & 11.20 & 9.8 & 9.8 \\ 
001.5$+$01.5 & JaSt46 & 13.24 & 0.03 & 10.73 & 4.5 & 4.4 \\ 
001.6$-$01.0 & JaSt90 & ND$^b$ & -- & -- & -- & -- \\ 
001.6$-$01.1 & JaSt97 & 14.54 & 0.09 & -- & 7.4 & 5.6 \\ 
001.6$-$01.3 & BlQ & 12.51 & 0.03 & 9.12 & 3.8 & 3.6 \\ 
001.6$-$01.5 & JaSt2-16 & 13.94 & 0.09 & 13.22 & 6.7 & 6.1 \\ 
001.6$-$02.6 & PHR1759-2853 & 12.75 & 0.03 & 11.12 & 22.6 & 21.3 \\ 
001.6$-$03.7 & MPA1804-2926 & 12.97$^a$ & 0.03 & 12.35 & 34.5 & 24.3 \\ 
001.6$+$01.5 & JaSt42 & 13.44 & 0.03 & 10.81 & 6.7 & 6.1 \\ 
001.6$+$01.6 & MPA1743-2636 & ND$^b$ & -- & -- & --  \\ 
001.7$-$01.6 & H2-31 & ND$^b$ & -- & -- & -- & -- \\ 
001.7$-$02.6 & PPA1800-2846 & 12.00 & 0.02 & 10.91 & 15.1 & 12.6 \\ 
001.8$-$03.8 & ShWi7 & 12.25 & 0.02 & 11.46 & 16.0 & 2.0 \\ 
001.7$-$04.4 & H1-55 & 13.31 & 0.03 & -- & 13.8 & 13.6 \\ 
001.7$-$04.6 & H1-56 & 11.33 & 0.02 & -- & 4.2 & 4.2 \\ 
001.8$-$01.5 & JaSt2-18 & ND$^b$ & -- & -- & --  & -- \\ 
001.8$-$02.0 & PHR1757-2824 & 13.13 & 0.03 & 10.44 & 19.3 & 8.0 \\ 
001.8$-$02.7 & PHR1800-2842 & 13.12 & 0.03 & 11.98 & 11.3 & 10.1 \\ 
001.8$-$03.2 & MPA1802-2900 & 12.85 & 0.03 & 12.05 & 22.4 & 21.4 \\ 
001.8$-$03.7 & PHR1804-2913 & 12.91 & 0.03 & 12.15 & 8.3 & 7.3 \\ 
001.9$-$02.5 & PPA1759-2834 & 13.24 & 0.03 & 11.68 & 15.6 & 13.5 \\ 
001.9$-$03.1 & MPA1802-2850 & ND$^a$$^b$ & -- & -- & -- & -- \\ 
002.0$-$01.3 & JaSt98 & 13.69 & 0.09 & 9.33 & 2.0 & 1.7 \\ 
\hline
    \end{tabular}
  \end{minipage}
\begin{minipage}{4.5in}  
\begin{tabular}{|l l lllll|}
\hline
 & & \multicolumn{3}{|c|}{$-\log$(F[OIII])} &  \multicolumn{2}{|c|}{Diameter ($\arcsec$)}\\ 
PN G name & Usual name & Flux & error & F$_{dered}^d$ & maj & min \\
(1) & (2) & (3) & (4) & (5) & (6) & (7) \\
\hline 
002.0$-$02.0  & H1-45 & 11.96 & 0.02 & -- & 1.5 & 1.1 \\ 
002.0$-$02.1 & MPA1758-2816 & ND$^a$$^b$ & -- & -- & -- & -- \\ 
002.0$-$02.4 & MPA1800-2825 & 13.52 & 0.09 & 12.19 & 14.2 & 12.5 \\ 
002.0$-$02.5 & PPA1800-2826 & 12.71 & 0.03 & 11.50 & 21.8 & 18.8 \\ 
002.0$-$03.1 & PHR1802-2847 & 12.77 & 0.03 & 11.69 & 19.5 & 18.2 \\ 
002.0$-$03.2 & PHR1803-2848 & 12.56 & 0.03 & 12.70 & 32.2 & 28.4 \\ 
002.0$-$03.4 & PPA1803-2855 & 13.11 & 0.03 & 12.53 & 19.1 & 18.6 \\ 
002.0$+$01.5 & PHR1744-2624 & 14.45 & 0.11& -- & 11.1 & 8.7 \\ 
002.1$-$01.1 & MPA1755-2741 & 14.66 & 0.10 & -- & 8.8 & 8.2 \\ 
002.1$-$02.2 & M3-20 & 11.21 & 0.02 & -- & 4.0 & 3.2 \\ 
002.1$-$02.4 & PPA1800-2818 & 12.53 & 0.03 & -- & 13.2 & 11.3 \\ 
002.1$-$02.8 & PHR1801-2831 & 12.51 & 0.03 & 11.16 & 18.5 & 16.4 \\ 
002.1$-$03.8 & MPA1805-2902 & 13.17$^a$ & 0.03 & 12.15 & 29.8 & 26.9 \\ 
002.1$-$04.1 & PHR1806-2909 & 12.87 & 0.03 & 12.35 & 28.1 & 27.7 \\ 
002.1$-$04.2 & H1-54 & 11.10 & 0.02 & -- & 1.9 & 1.6 \\ 
002.1$+$01.2 & MPA1745-2626 & ND$^b$ & -- & -- & -- & -- \\ 
002.1$+$03.3 & & 13.66 & 0.09 & -- & 5.8 & 5.2 \\ 
002.2$-$01.2 & PPA1755-2739 & 13.73 & 0.09 & -- & 7.1 & 5.8 \\ 
002.2$-$02.5 & KFL2 & 12.38 & 0.02 & 11.14 & 8.2 & 5.9 \\ 
002.2$-$04.3 & PHR1808-2913 & 12.79 & 0.03 & -- & 2.2 & 2.1 \\ 
002.3$-$01.7 & PPA1757-2750 & ND$^b$ & -- & -- & -- & -- \\ 
002.3$-$01.9 & PHR1758-2756 & 13.33 & 0.03 & -- & 19.9 & 18.3 \\ 
002.3$-$03.4 & H2-37 & 11.78 & 0.02 & 10.69 & 6.0 & 3.5 \\ 
002.3$+$01.7 & PHR1744-2603 & 14.00 & 0.09 & 11.47 & 8.8 & 8.1 \\ 
002.3$+$02.2 & K5-11 & 13.01 & 0.03 & 10.79 & 10.2 & 9.1 \\ 
002.3$+$02.4 & PPA1741-2538 & 13.13 & 0.03 & 11.20 & 8.5 & 6.8 \\ 
002.3$+$03.6 & PPA1737-2501 & 13.12 & 0.03 & -- & 6.4 & 6.2 \\ 
002.4$-$02.6 & PHR1801-2809 & 14.05 & 0.09 & 12.64 & 32.6 & 29.4 \\ 
002.4$-$03.1 & PPA1803-2826 & 12.00 & 0.02 & 10.78 & 4.1 & 3.2 \\ 
002.4$-$03.2 & Sa3-115 & 11.92 & 0.02 & 10.89 & 18.6 & 15.4 \\ 
002.4$-$03.4 & PHR1804-2833 & 12.56 & 0.03 & -- & 1.7 & 0.4 \\ 
002.4$-$03.7 & M1-38 & 13.21 & 0.03 & 12.78 & 3.7 & 2.6 \\ 
002.4$+$02.1 & PHR1743-2541 & 13.83 & 0.09 & 11.67 & 14.9 & 11.2 \\ 
002.4$+$03.5 & PHR1738-2500 & 13.66 & 0.09 & -- & 15.7 & 14.3 \\ 
002.5$-$01.5 & MPA1757-3532 & 14.53$^a$ & 0.10 & -- & 14.7 & 12.5 \\ 
002.5$-$01.7 & Pe2-11 & 12.86 & 0.03 & 10.66 & 7.2 & 6.5 \\ 
002.5$-$02.6 & MPA1802-2803 & 13.57$^a$ & 0.09 & 12.64 & 42.8 & 41.9 \\ 
002.5$+$01.9 & PHR1744-2545 & 13.86 & 0.09 & 11.40 & 15.7 & 13.0 \\ 
002.5$+$02.0 & PHR1743-2538 & 13.56 & 0.09 & 10.66 & 5.1 & 4.4 \\ 
002.6$-$01.7 & PHR1758-2729 & 13.38 & 0.03 & 11.00 & 8.7 & 6.5 \\ 
002.6$-$02.5 & MPA1801-2755 & 13.49$^a$ & 0.03 & 11.66 & 12.1 & 9.4 \\ 
002.6$-$02.8 & PHR1803-2804 & 13.66 & 0.09 & 12.21 & 39.2 & 17.2 \\ 
002.6$-$02.8a & MPA1802-2807 & 13.13$^a$ & 0.03 & 11.68 & 40.3 & 37.4 \\ 
002.6$-$03.1 & PHR1804-2816 & 12.50 & 0.03 & 11.33 & 14.3 & 14.0 \\ 
002.6$-$03.4 & M1-37 & 13.41 & 0.03 & -- & 0.8 & 0.7 \\ 
002.6$+$02.1 & K5-13 & 12.45 & 0.02 & 10.40 & 11.7 & 9.9 \\ 
\hline
    \end{tabular}
  \end{minipage}
\end{table}
\end{landscape}
%------------------------------------------------------------------------------------------------
\begin{landscape}
\begin{table}
\contcaption{ }
\begin{minipage}{4.5in}
 \begin{tabular}{|l l l l l l l|}
\hline
 & & \multicolumn{3}{|c|}{$-\log$(F[OIII])} &  \multicolumn{2}{|c|}{Diameter ($\arcsec$)}\\ 
PN G name & Usual name & Flux & error & F$_{dered}^d$ & maj & min \\
(1) & (2) & (3) & (4) & (5) & (6) & (7) \\
\hline 
002.6$+$02.3 & PHR1742-2525 & 13.42 & 0.03 & 11.02 & 11.5 & 11.4 \\
002.7$-$02.4 & PPA1801-2746 & 14.01 & 0.09 & 12.46 & 11.5 & 8.5 \\ 
002.7$-$03.7 & PHR1806-2824 & 12.80 & 0.03 & 11.77 & 13.1 & 12.6 \\ 
002.7$-$04.6 & MPA1810-2851 & 15.01$^a$ & 0.18 & -- & 3.0 & 1.4 \\ 
002.7$-$04.8 & M1-42 & 10.92 & 0.02 & -- & 12.5 & 11.6 \\ 
002.8$+$01.7 & H2-20 & 13.79 & 0.09 & 11.59 & 2.8 & 2.7 \\ 
002.7$+$01.7 & PPA1745-2542 & 13.90 & 0.09 & 10.91 & 3.6 & 3.5 \\ 
002.8$-$02.2 & Pe2-12 & 13.88 & 0.09 & 12.34 & 8.3 & 1.8 \\ 
002.8$-$04.1 & PHR1808-2835 & 12.15 & 0.02 & 10.73 & 18.9 & 16.3 \\ 
002.8$+$01.7 & K5-16 & 12.76 & 0.03 & 10.31 & 11.8 & 8.9 \\ 
002.9$-$01.8 & MPA1759-2719 & 13.73$^a$ & 0.09 & -- & 20.6 & 20.6 \\ 
002.9$-$03.0 & PHR1804-2757 & 13.03 & 0.03 & 11.84 & 47.3 & 40.4 \\ 
002.9$-$03.3 & PHR1805-2804 & ND$^b$ & -- & -- & -- & -- \\ 
002.9$-$03.6 & MPA1806-2812 & ND$^a$$^b$ & -- & -- & -- & -- \\ 
002.9$-$04.0 & H2-39 & 11.68 & 0.02 & -- & 6.9 & 4.7 \\ 
003.0$-$03.6 & MPA1806-2807 & 12.89$^a$ & 0.03 & -- & 19.6 & 17.3 \\ 
003.1$-$01.6 & PHR1759-2706 & 13.47 & 0.03 & 11.10 & 30.5 & 29.9 \\ 
003.1$-$02.1 & PHR1801-2718 & 13.21 & 0.03 & 11.41 & 38.1 & 35.1 \\ 
003.1$+$02.9 & Hb4 & 10.79 & 0.02 & -- & 9.8 & 7.5 \\ 
003.1$+$03.4 & H2-17 & 14.36 & 0.09 & -- & 2.3 & 1.9 \\ 
003.2$-$04.4 & KFL12 & 12.24 & 0.02 & -- & 3.4 & 3.1 \\ 
003.3$-$04.4 & PPA1810-2813 & 12.73 & 0.03 & -- & 13.2 & 13.0 \\ 
003.3$-$04.6 & Ap1-12 & 12.52 & 0.03 & -- &  &  \\ 
003.4$-$01.8 & PHR1800-2653 & 13.07 & 0.03 & 10.54 & 9.3 & 9.5 \\ 
003.4$-$04.8 & H2-43 & 12.46 & 0.02 & -- & 1.1 & 0.9 \\ 
003.4$+$04.3 & PHR1737-2341 & 13.59 & 0.09 & -- & 17.2 & 14.0 \\ 
003.5$-$02.3 & PHR1803-2702 & ND$^b$ & -- & -- & -- & -- \\ 
003.5$-$02.4 & IC4673 & 10.69 & 0.02 & 9.71 & 21.7 & 16.5 \\ 
003.5$-$02.9 & MPA1805-2721 & 13.10 & 0.03 & 12.25 & 31.1 & 22.2 \\ 
003.5$-$04.6 & NGC6565 & 10.14 & 0.02 & -- & 10.5 & 9.6 \\ 
003.5$+$02.6 & PHR1743-2431 & 12.56 & 0.03 & -- & 33.9 & 33.6 \\
003.5$+$04.3 & PPA1737-2341 & ND$^b$ & -- & -- & -- & -- \\ 
003.5$+$04.5 & PHR1736-2330 & 13.50 & 0.09 & -- & 10.8 & 5.0 \\ 
003.6$-$02.3 & M2-26 & 11.91 & 0.02 & 10.56 & 9.5 & 9.4 \\ 
003.6$-$02.8 & MPA1805-2712 & 13.21 & 0.03 & 11.76 & 42.9 & 41.1 \\ 
003.6$-$03.0 & PHR1805-2723 & 12.34 & 0.02 & 11.26 & 21.6 & 19.9 \\ 
003.6$+$02.7 & PHR1743-2424 & 13.99 & 0.09 & -- & 22.8 & 21.8 \\ 
003.6$+$03.1 & M2-14 & 11.85 & 0.02 & -- & 3.0 & 2.6 \\ 
003.6$+$04.9 & K5-6 & 12.41 & 0.02 & -- & 22.6 & 9.7 \\ 
003.7$-$04.6 & M2-30 & 10.76 & 0.02 & -- & 5.1 & 5.0 \\ 
003.8$-$02.4 & PHR1804-2653 & 13.40 & 0.03 & -- & 8.8 & 6.7 \\ 
003.8$-$03.2 & PHR1807-2715 & 14.06 & 0.09 & -- & 12.4 & 10.0 \\ 
003.8$-$04.3 & H1-59 & 11.44 & 0.02 & -- & 6.6 & 6.0 \\ 
003.8$-$04.5 & H2-41 & 11.79 & 0.02 & -- & 10.5 & 9.6 \\ 
003.9$+$02.6 & K5-14 & 12.16 & 0.02 & -- & 1.8 & 1.7 \\
003.9$-$02.3 & M1-35 & 11.33 & 0.02 & -- & 7.3 & 6.8 \\
\hline
    \end{tabular}
  \end{minipage}
\begin{minipage}{4.5in}  
\begin{tabular}{|l l lllll|}
\hline
 & & \multicolumn{3}{|c|}{$-\log$(F[OIII])} &  \multicolumn{2}{|c|}{Diameter ($\arcsec$)}\\ 
PN G name & Usual name & Flux & error & F$_{dered}^d$ & maj & min \\
(1) & (2) & (3) & (4) & (5) & (6) & (7) \\
\hline   
003.9$-$03.1 & KFL7 & 12.34 & 0.02 & -- & 8.1 & 5.0 \\ 
003.9$+$01.6 & & 13.60 & 0.09 & -- & 9.7 & 9.5 \\ 
004.0$-$02.5 & PHR1804-2642 & 13.41 & 0.03 & 11.45 & 15.0 & 12.6 \\ 
004.0$-$02.6 & PHR1804-2645 & 12.40 & 0.02 & 10.40 & 24.5 & 13.2 \\ 
004.0$-$02.7 & PPA1805-2649 & 13.11 & 0.03 & 11.46 & 27.4 & 22.1 \\ 
004.0$-$03.0 & M2-29 & 11.45 & 0.02 & -- & 4.8 & 3.6 \\ 
004.1$-$03.3 & PPA1808-2700 & 12.38 & 0.02 & -- & 13.1 & 8.5 \\ 
004.1$-$03.8 & KFL11 & 12.26 & 0.02 & -- & 3.0 & 2.3 \\ 
004.2$-$04.3 & H1-60 & 11.38 & 0.02 & -- & 6.1 & 5.6 \\ 
004.1$+$01.7 & PPA1748-2427 & 13.43 & 0.03 & -- & 25.7 & 16.9 \\ 
004.2$-$02.5 & PHR1805-2631 & 12.48 & 0.02 & 10.41 & 12.5 & 10.3 \\ 
004.2$-$03.2 & KFL10 & 11.93 & 0.02 & -- & 7.1 & 5.6 \\ 
004.2$+$01.5 & K6-29 & 13.18 & 0.03 & -- & 10.8 & 7.7 \\ 
004.2$+$02.0 & K5-17 & 12.10 & 0.02 & -- & 4.6 & 3.1 \\ 
004.2$+$02.0a & MPA1747-2414 & 13.72 & 0.09 & -- & 25.2 & 23.1 \\ 
004.3$-$02.6 & H1-53 & 12.36 & 0.02 & 10.46 & 2.3 & 1.7 \\ 
004.3$+$01.8 & H2-24 & 12.50 & 0.03 & -- & 8.4 & 4.3 \\ 
004.3$+$01.8a & PHR1748-2417 & 12.81 & 0.03 & -- & 16.1 & 14.8 \\ 
004.4$-$03.1 & PHR1807-2637 & 13.15 & 0.03 & -- & 25.4 & 25.3 \\ 
004.5$-$03.0 & MPA1807-2631 & 13.17$^a$ & 0.03 & -- & 41.4 & 40.4 \\ 
004.5$+$02.0 & MPA1748-2402 & 13.27 & 0.03 & -- & 12.9 & 9.4 \\ 
004.6$+$01.8 & BMP1749-2356 & 13.63 & 0.09 & -- & 26.1 & 17.9 \\ 
004.8$-$00.5 & PHR1759-2501 & 13.04 & 0.03 & -- & 42.3 & 26.4 \\ 
004.8$-$01.1 & PHR1801-2522 & 14.16 & 0.09 & -- & 7.3 & 7.0 \\ 
004.8$+$02.0 & H2-25 & 13.41 & 0.03 & -- & 3.1 & 3.0 \\ 
355.0$-$03.3 & PPA1746-3454 & 14.25 & 0.09 & 11.75 & 8.3 & 2.9 \\ 
355.0$-$03.7 & K5-18 & 12.28 & 0.02 & 11.04 & 11.1 & 10.0 \\ 
355.1$-$02.9 & H1-31 & 11.40 & 0.02 & 9.74 & 1.8 & 1.7 \\ 
355.1$+$03.7 & PHR1718-3055 & 13.09 & 0.03 & 11.58 & 20.8 & 19.9 \\ 
355.2$-$02.0 & PPA1741-3405 & 13.35 & 0.03 & -- & 6.5 & 6.5 \\ 
355.2$-$02.5 & H1-29 & 11.61 & 0.02 & 10.25 & 2.4 & 2.2 \\ 
355.2$-$05.0 & PHR1754-3533 & 12.99 & 0.03 & -- & 26.9 & 26.1 \\ 
355.2$+$03.7 & Th3-5 & 12.62 & 0.03 & 11.09 & 13.3 & 10.5 \\ 
355.3$-$03.2 & PPA1747-3435 & 13.34 & 0.03 & 11.98 & 19.5 & 15.4 \\ 
355.3$-$04.1 & PHR1750-3500 & 12.65 & 0.03 & -- & 20.4 & 16.7 \\ 
355.3$+$03.7 & MPA1719-3043 & 12.67 & 0.03 & -- & 72.6 & 72.2 \\ 
355.4$-$01.4 & K6-9 & 13.44 & 0.03 & -- & 8.4 & 7.5 \\ 
355.4$-$02.4 & M3-14 & 11.45 & 0.02 & 10.05 & 3.8 & 3.6 \\ 
355.4$-$02.6 & PHR1745-3413 & 13.10 & 0.03 & 11.70 & 15.2 & 11.9 \\ 
355.4$-$03.1 & PPA1746-3428 & 12.97 & 0.03 & -- & 26.8 & 24.2 \\ 
355.4$-$04.0 & Pe1-10 & 10.92 & 0.02 & Pe1-10 & 18.2 & 14.0 \\ 
355.4$-$04.3 & K5-34 & 11.91 & 0.02 & -- & 6.8 & 5.8 \\ 
355.4$+$03.6 & PHR1719-3044 & 13.34 & 0.03 & 11.65 & 20.0 & 15.6 \\ 
355.5$-$01.1 & MPA1739-3320 & 14.81 & 0.24 & -- & 7.2 & 6.0 \\ 
355.5$-$03.7 & PHR1749-3438 & 12.31 & 0.02 & 11.00 & 19.6 & 13.2 \\ 
355.5$-$04.8 & PHR1754-3515 & 12.54 & 0.03 & -- & 19.5 & 18.7 \\ 
 
\hline
    \end{tabular}
  \end{minipage}
\end{table}
\end{landscape}
%------------------------------------------------------------------------------------------------
\begin{landscape}
\begin{table}
\contcaption{ }
\begin{minipage}{4.5in}
 \begin{tabular}{|l l l l l l l|}
\hline
 & & \multicolumn{3}{|c|}{$-\log$(F[OIII])} &  \multicolumn{2}{|c|}{Diameter ($\arcsec$)}\\ 
PN G name & Usual name & Flux & error & F$_{dered}^d$ & maj & min \\
(1) & (2) & (3) & (4) & (5) & (6) & (7) \\
\hline
355.5$+$03.6 & PHR1720-3041 & 13.78 & 0.09 & -- & 24.6 & 22.3 \\ 
355.6$-$01.4 & PHR1740-3324 & 13.52 & 0.09 & -- & 11.8 & 9.6 \\ 
355.6$-$02.3 & PHR1744-3355 & 12.16 & 0.02 & 11.04 & 65.5 & 63.7 \\ 
355.6$+$03.6 & PHR1720-3038 & 13.07 & 0.03 & 11.11 & 15.2 & 14.9 \\ 
355.6$+$04.1 & PHR1718-3019 & 12.60 & 0.03 & -- & 46.7 & 40.5 \\ 
355.7$-$03.0 & H1-33 & 11.63 & 0.02 & 10.40 & 4.0 & 3.2 \\ 
355.7$-$03.4 & H2-23 & 11.66 & 0.02 & 10.41 & 3.4 & 2.7 \\ 
355.8$+$03.5 & PHR1721-3027 & 13.06 & 0.03 & 10.78 & 13.5 & 8.8 \\ 
355.8$+$04.5 & PHR1717-2954 & 13.69 & 0.09 & -- & 12.2 & 11.2 \\ 
355.9$-$04.2 & M1-30 & 11.54 & 0.02 & -- & 4.3 & 3.6 \\ 
355.9$+$02.7 & Th3-10 & 13.08 & 0.03 & 9.96 & 3.0 & 2.6 \\ 
355.9$+$03.1 & PPA1723-3038 & 13.76 & 0.09 & 10.42 & 5.6 & 5.1 \\ 
355.9$+$04.1 & PHR1719-3003 & 13.28 & 0.03 & -- & 23.5 & 23.0 \\ 
356.0$-$01.4 & PPA1741-3302 & 14.01 & 0.09 & -- & 6.9 & 6.8 \\ 
356.0$-$01.8 & PPA1743-3315 & 13.77 & 0.09 & -- & 4.9 & 4.4 \\ 
356.0$-$04.2 & PHR1753-3428 & 12.46 & 0.02 & 11.09 & 15.3 & 14.6 \\ 
356.0$+$02.8 & PPA1724-3043 & 13.97 & 0.09 & 10.71 & 10.6 & 9.3 \\ 
356.1$-$02.1 & PHR1744-3319 & 13.68 & 0.09 & -- & 8.9 & 8.4 \\ 
356.1$-$02.7 & PPA1747-3341 & 13.97 & 0.09 & 11.17 & 7.3 & 7.0 \\ 
356.1$-$03.3 & H2-26 & 12.50 & 0.03 & 11.14 & 5.5 & 5.0 \\ 
356.1$+$02.7 & Th3-13 & 12.82 & 0.03 & 10.00 & 1.9 & 1.4 \\ 
356.2$-$03.2 & PHR1749-3347 & 13.58 & 0.09 & 12.16 & 12.8 & 12.5 \\ 
356.2$-$03.6 & PPA1751-3401 & 12.58 & 0.03 & 11.20 & 11.4 & 11.2 \\ 
356.2$-$04.4 & Cn2-1 & 10.38 & 0.02 & -- & 2.6 & 2.6 \\ 
356.2$+$02.5 & PPA1726-3045 & 14.17 & 0.09 & 9.97 & 4.1 & 2.7 \\ 
356.3$-$02.6 & MPA1747-3326 & 12.74 & 0.03 & 11.29 & 8.1 & 7.3 \\ 
356.4$-$02.5 & K6-12 & 13.39 & 0.03 & 11.43 & 13.7 & 11.2 \\ 
356.5$-$01.8 & PPA1744-3252 & 13.82 & 0.09 & -- & 4.6 & 4.4 \\ 
356.5$-$02.3 & M1-27 & ND$^b$ & -- & -- & -- & -- \\ 
356.5$-$03.4 & MPA1751-3339 & 13.11 & 0.03 & 11.23 & 8.1 & 8.0 \\
356.5$-$03.6 & H2-27 & 12.64 & 0.03 & 10.53 & 5.2 & 4.2 \\ 
356.5$-$03.9 & H1-39 & 12.92 & 0.03 & -- & 2.3 & 2.0 \\ 
356.5$-$04.1 & PPA1754-3358 & 12.72 & 0.03 & 11.43 & 14.1 & 13.3 \\ 
356.6$-$01.9 & PHR1745-3246 & 12.61 & 0.03 & -- & 51.6 & 34.4 \\ 
356.6$-$04.7 & PHR1756-3414 & 12.29 & 0.02 & 11.47 & 20.1 & 18.7 \\ 
356.7$-$04.7 & MPA1757-3410 & 12.79 & 0.03 & 11.85 & 12.0 & 11.3 \\ 
356.7$-$04.8 & H1-41 & 10.89 & 0.02 & 10.40 & 12.0 & 8.8 \\ 
356.8$-$03.0 & K5-20 & 12.36 & 0.02 & 10.70 & 5.3 & 5.2 \\ 
356.8$-$03.6 & PHR1752-3330 & 13.42 & 0.03 & 12.57 & 1.1 & 0.9 \\ 
356.8$+$03.3 & Th3-12 & 14.32 & 0.09 & -- & 2.0 & 1.3 \\ 
356.9$+$02.2 & MPA1729-3016 & 13.50 & 0.09 & 11.69 & 7.0 & 6.9 \\ 
356.9$+$04.4 & M3-38 & 11.78 & 0.02 & -- & 1.6 & 1.2 \\ 
357.0$-$04.4 & PHR1756-3342 & 13.16 & 0.03 & 11.57 & 21.6 & 20.9 \\ 
357.1$-$04.7 & H1-43 & ND$^b$ & -- & -- & -- & -- \\ 
357.1$+$01.9 & Th3-24 & 13.23 & 0.03 & 11.74 & 8.6 & 7.3 \\ 
357.1$+$03.6 & M3-7 & 11.61 & 0.02 & -- & 6.5 & 6.0 \\ 
\hline
    \end{tabular}
  \end{minipage}
\begin{minipage}{4.5in}  
\begin{tabular}{|l l lllll|}
\hline
 & & \multicolumn{3}{|c|}{$-\log$(F[OIII])} &  \multicolumn{2}{|c|}{Diameter ($\arcsec$)}\\ 
PN G name & Usual name & Flux & error & F$_{dered}^d$ & maj & min \\
(1) & (2) & (3) & (4) & (5) & (6) & (7) \\
\hline 
357.1$+$04.4 & PBOZ1 & 12.51 & 0.03 & -- & 10.9 & 9.1 \\ 
357.2$-$04.5 & H1-42 & 10.54 & 0.02 & -- & 4.3 & 3.7 \\ 
357.2$+$01.4 & Al2-H & 12.90 & 0.03 & 10.22 & 8.5 & 7.0 \\ 
357.2$+$02.0 & H2-13 & 12.05 & 0.02 & 9.86 & 5.6 & 5.4 \\ 
357.3$-$02.0 & PPA1747-3215 & ND$^b$ & -- & -- & -- & -- \\ 
357.3$+$01.3 & PHR1733-3029 & 13.86 & 0.09 & -- & 17.4 & 17.1 \\ 
357.3$+$02.3 & K6-25 & 13.24 & 0.03 & 10.51 & 5.0 & 5.0 \\ 
357.3$+$03.3 & M3-41 & 14.06 & 0.09 & -- & HS$^c$ & HS$^c$  \\ 
357.3$+$04.0 & H2-7 & 11.95 & 0.02 & -- & 4.8 & 4.3 \\ 
357.3$+$04.1 & PHR1723-2856 & 13.38 & 0.03 & -- & 20.7 & 18.9 \\ 
357.4$-$03.1 & PHR1752-3244 & 12.85 & 0.03 & -- & 10.8 & 10.6 \\ 
357.4$-$03.2 & M2-16 & 11.36 & 0.02 & 10.35 & 3.1 & 2.9 \\ 
357.4$-$03.5 & M2-18 & 11.67 & 0.02 & -- & 2.2 & 2.1 \\ 
357.4$-$04.6 & M2-22 & 11.54 & 0.02 & 10.42 & 5.8 & 5.2 \\ 
357.4$+$03.4 & PPA1725-2915 & 12.80 & 0.03 & -- & 11.2 & 10.7 \\ 
357.5$-$02.4 & PPA1749-3216 & 13.68 & 0.09 & -- & 7.8 & 6.5 \\ 
357.5$+$01.3 & PPA1734-3015 & ND$^b$ & -- & -- & -- & -- \\ 
357.5$+$02.0 & PPA1731-2955 & 13.55 & 0.09 & -- & 19.0 & 12.4 \\ 
357.6$-$03.0 & PHR1752-3233 & 12.64 & 0.03 & -- & 10.9 & 7.5 \\ 
357.6$-$03.0a & PHR1752-3230 & 12.98 & 0.03 & -- & 9.7 & 9.1 \\ 
357.6$-$03.3 & H2-29 & 13.29 & 0.03 & 11.50 & 9.0 & 8.0 \\ 
357.6$+$01.0 & TBJ4 & 12.98 & 0.03 & 10.64 & 31.5 & 26.8 \\ 
357.6$+$01.7 & H1-23 & 11.84 & 0.02 & 9.70 & 3.5 & 2.6 \\ 
357.7$-$01.7 & & ND$^b$ & -- & --& -- & -- \\ 
357.7$-$04.8 & BMP1759-3321 & ND$^b$ & -- & -- & -- & -- \\ 
357.7$+$01.4 & PPA1734-3004 & 13.81 & 0.09 & -- & 3.6 & 3.1 \\ 
357.8$-$03.3 & PHR1753-3228 & 11.75 & 0.02 & 10.32 & 65.1 & 40.1 \\ 
357.8$-$04.4 & PHR1758-3304 & 11.98 & 0.02 & 10.69 & 16.6 & 14.3 \\ 
357.8$+$01.6 & PPA1734-2954 & 14.01 & 0.09 & -- & 12.7 & 8.6 \\ 
357.9$+$01.7 & PPA1733-2945 & 13.57 & 0.09 & 11.10 & 5.7 & 5.0 \\ 
358.0$-$02.4 & PPA1750-3152 & 13.66 & 0.09 & -- & 4.9 & 3.7 \\ 
358.0$-$02.7 & Al2-O & 12.29 & 0.02 & -- & 9.1 & 7.6 \\ 
358.0$-$04.6 & Sa3-107 & 12.42 & 0.02 & 10.82 & 6.3 & 5.9 \\ 
358.0$+$01.5 & JaSt1 & 14.28 & 0.09 & 11.87 & 7.1 & 5.1 \\ 
358.0$+$01.6 & PHR1734-2944 & 14.04 & 0.09 & 10.76 & 8.9 & 6.8 \\ 
358.0$+$02.6 & Th3-23 & 12.61 & 0.03 & 9.97 & 6.9 & 5.9 \\ 
358.1$+$02.3 & PPA1731-2915 & 14.80 & 0.20 & -- & 3.8 & 1.5 \\ 
358.2$-$01.1 & & 12.53 & 0.03 & -- & 17.5 & 15.7 \\ 
358.2$+$03.5 & H2-10 & 11.92 & 0.02 & 9.89 & 3.7 & 3.0 \\ 
358.2$+$03.6 & M3-10 & 11.13 & 0.02 & -- & 4.2 & 4.0 \\ 
358.3$-$02.5 & M4-7 & 12.29 & 0.02 & 9.65 & 6.9 & 6.6 \\ 
358.3$+$01.2 & BlB & 13.25 & 0.03 & -- & 2.4 & 2.0 \\ 
358.3$+$03.0 & H1-17 & 11.63 & 0.02 & 9.53 & 2.8 & 2.8 \\ 
358.4$-$02.3 & MPA1751-3128 & 14.19 & 0.09 & -- & 15.1 & 12.2 \\ 
358.4$+$01.6 & JaSt3 & 13.61 & 0.09 & 9.90 & 7.8 & 7.8 \\ 
358.4$+$01.7 & JaSt2 & 14.34 & 0.09 & 10.78 & 4.4 & 4.3 \\ 
\hline
    \end{tabular}
  \end{minipage}
\end{table}
\end{landscape}
%------------------------------------------------------------------------------------------------
\begin{landscape}
\begin{table}
\contcaption{ }
\begin{minipage}{4.5in}
 \begin{tabular}{|l l l l l l l|}
\hline
 & & \multicolumn{3}{|c|}{$-\log$(F[OIII])} &  \multicolumn{2}{|c|}{Diameter ($\arcsec$)}\\ 
PN G name & Usual name & Flux & error & F$_{dered}^d$ & maj & min \\
(1) & (2) & (3) & (4) & (5) & (6) & (7) \\
\hline
358.4$+$02.7 & PHR1731-2850 & 13.16 & 0.03 & -- & 22.1 & 16.1 \\ 
358.4$+$03.3 & Th3-19 & 12.27 & 0.02 & 9.96 & 2.3 & 2.3 \\ 
358.5$-$01.7 & JaSt61 & 13.90 & 0.09 & 10.69 & 2.0 & 2.0 \\ 
358.5$-$04.2 & H1-46 & 11.26 & 0.02 & -- & 1.5 & 1.2 \\ 
358.5$+$02.6 & K6-1 & 11.50 & 0.02 & -- & 47.0 & 34.2 \\ 
358.5$+$02.9 & Al2-F & 12.71 & 0.03 & 10.91 & 4.2 & 3.5 \\ 
358.5$+$03.7 & Al2-B & 13.02 & 0.03 & 11.28 & 6.8 & 4.6 \\ 
358.6$-$02.4 & K6-16 & 13.66 & 0.09 & 11.07 & 8.9 & 8.2 \\ 
358.6$-$03.6 & PHR1756-3157 & 11.96 &  0.02& -- & 53.1 & 46.5 \\ 
358.6$+$01.7 & JaSt4 & 13.96 & 0.09 & 10.78 & 10.6 & 9.5 \\ 
358.6$+$01.8 & M4-6 & 12.43 & 0.02 & 9.80 & 4.1 & 4.0 \\ 
358.7$-$02.5 & PHR1752-3116 & 13.04 & 0.03 & 10.90 & 9.3 & 6.9 \\ 
358.7$-$02.7 & Al2-R & 12.93 & 0.03 & 11.00 & 6.4 & 3.9 \\ 
358.7$-$03.0 & K6-34 & 13.53 & 0.09 & 11.80 & 10.4 & 9.8 \\ 
358.8$+$01.7 & JaSt5 & 13.44 & 0.03 & 10.64 & 9.1 & 5.9 \\ 
358.8$+$03.0 & Th3-26 & 12.15 & 0.02 & 10.53 & 9.1 & 8.3 \\ 
358.8$+$03.4 & MPA1729-2804 & 13.96 & 0.09 & -- & 14.8 & 14.6 \\ 
358.8$+$03.8 & PHR1727-2747 & 13.56 & 0.09 & 11.42 & 16.2 & 14.4 \\ 
358.9$-$01.5 & JaSt65 & 14.30 & 0.09 & 10.36 & 5.7 & 5.6 \\ 
358.9$-$02.1 & PHR1751-3059 & 13.64 & 0.09 & 12.02 & 13.8 & 13.5 \\ 
358.9$-$03.6 & PPA1757-3144 & 12.85 & 0.03 & -- & 13.8 & 6.3 \\ 
358.9$-$03.7 & H1-44 & 13.03 & 0.03& --  & 3.5 & 3.3 \\ 
358.9$+$03.2 & H1-20 & 11.81 & 0.02 & 9.74 & 4.4 & 3.8 \\ 
358.9$+$03.4 & H1-19 & 12.32 & 0.02 & 10.52 & 2.6 & 2.0 \\ 
359.0$-$01.6 & GLMP647 & ND$^b$ & -- & -- & -- & -- \\ 
359.0$-$04.1 & M3-48 & 12.39 & 0.02 & -- & 4.7 & 4.2 \\ 
359.0$-$04.8 & M2-25 & 11.45 & 0.02 & -- & 17.7 & 13.4 \\ 
359.0$-$04.9 & PHR1803-3218 & 12.69 & 0.03 & -- & 35.1 & 30.2 \\ 
359.0$+$01.1 & JaSt9 & 14.71 & 0.22 & 13.62 & 6.4 & 2.4 \\ 
359.0$+$02.8 & Al2-G & 13.15 & 0.03 & 10.82  & 3.5 & 1.0 \\ 
359.0$+$03.0 & MPA1731-2805 & 14.18 & 0.10 & -- & 20.7 & 12.4 \\ 
359.0$+$03.7 & PHR1728-2743 & 13.40 & 0.03 & 10.75 & 18.4 & 17.6 \\ 
359.1$-$01.7 & M1-29 & 11.13 & 0.02 & 9.50 & 9.1 & 6.7 \\ 
359.1$-$02.3 & M3-16 & 11.66 & 0.02 & 10.09 & 10.0 & 7.7 \\ 
359.1$+$02.9 & MPA1732-2806 & 14.44 & 0.10 & -- & 16.5 & 6.0 \\ 
359.2$-$03.1 & PHR1756-3112 & 13.00 & 0.03 & 11.45 & 16.6 & 13.4 \\ 
359.2$+$01.2 & & 13.66 & 0.09 & 9.13 & 23.0 & 4.7 \\ 
359.2$+$01.3 & JaSt8 & 14.10 & 0.09 & -- & 8.0 & 6.7 \\ 
359.3$-$00.9 & Hb5 & 10.33 & 0.02 & -- & 5.6 & 5.5 \\ 
359.3$-$02.3 & PHR1753-3038 & 13.73 & 0.09 & -- & 14.1 & 11.3 \\ 
359.3$-$02.8 & MPA1755-3058 & 13.61 & 0.09 & -- & 37.0 & 33.8 \\ 
359.3$-$03.1 & M3-17 & 12.83 & 0.03 & 11.38 & 3.0 & 2.9 \\ 
359.3$+$01.3 & JaSt12 & 14.53 & 0.13 & -- & 3.9 & 2.4 \\ 
359.3$+$01.4 & & 13.40 & 0.03 & 9.69 & 3.3 & 2.6 \\ 
359.3$+$01.4a & PHR1738-2847 & 14.45 & 0.09 & -- & 6.9 & 6.3 \\ 
359.3$+$03.6 & Al2-E & 12.41 & 0.02 & 9.96 & 8.8 & 8.0 \\ 
\hline
    \end{tabular}
  \end{minipage}
\begin{minipage}{4.5in}  
\begin{tabular}{|l l lllll|}
\hline
 & & \multicolumn{3}{|c|}{$-\log$(F[OIII])} &  \multicolumn{2}{|c|}{Diameter ($\arcsec$)}\\ 
PN G name & Usual name & Flux & error & F$_{dered}^d$ & maj & min \\
(1) & (2) & (3) & (4) & (5) & (6) & (7) \\
\hline 
359.4$-$03.3 & PHR1757-3106 & 12.84 & 0.03 & 12.18 & 12.8 & 10.4 \\ 
359.4$-$03.4 & H2-33 & 12.11 & 0.02 & 10.60 & 7.9 & 7.3 \\ 
359.4$+$02.3 & Th3-32 & ND$^b$ & -- & -- & -- & -- \\ 
359.4$+$02.3a & PPA1735-2809 & 13.93 & 0.09 & 10.33 & 4.5 & 2.9 \\ 
359.4$-$03.7 & MPA1759-3116 & 13.51 & 0.09 & 12.59 & 22.9 & 22.6 \\ 
359.5$-$01.2 & JaSt66 & 14.14 & 0.09 & 9.39 & 3.4 & 2.7 \\ 
359.5$-$01.3 & JaSt68 & 15.06 & 0.19 & -- & 2.0 & 1.7 \\ 
359.5$-$01.8 & PHR1751-3012 & 14.64 & 0.11 & 12.64 & 38.0 & 32.1 \\ 
359.5$+$02.6 & Al2-K & 12.44 & 0.02 & 9.94 & 5.6 & 4.8 \\ 
359.5$+$03.5 & MPA1730-2726 & 13.71 & 0.09 & -- & 20.0 & 18.5 \\ 
359.6$-$04.8 & H2-36 & 11.85 & 0.02 & -- & 17.7 & 14.5 \\ 
359.6$+$01.0 & & 14.69 & 0.09 & 10.01 & 3.0 & 2.4 \\ 
359.6$+$02.2 & & 13.40 & 0.03 & 10.06 & 4.7 & 4.5 \\ 
359.7$-$01.4 & JaSt73 & 15.06 & 0.19 & 12.19 & 1.2 & 0.7 \\ 
359.7$-$01.7 & K6-15 & 13.38 & 0.03 & 10.70 & 5.0 & 4.4 \\ 
359.7$-$01.8 & M3-45 & 11.89 & 0.02 & 9.69 & 7.1 & 6.5 \\ 
359.7$-$02.2 & PPA1753-3021 & 13.68 & 0.09 & 11.80 &  18.5 & 18.3 \\ 
359.7$-$02.6 & H1-40 & 11.69 & 0.02 & -- & 1.4 & 1.4 \\ 
359.7$-$04.4 & KFL3 & 12.16 & 0.02 & 10.91 & 15.2 & 14.3 \\ 
359.7$-$04.4a & PPA1802-3124 & 12.49 & 0.02 & 11.70 & 17.6 & 15.5 \\ 
359.7$+$02.0 & PPA1736-2804 & 13.55 & 0.09 & 9.60 & 4.3 & 3.6 \\ 
359.8$+$01.0 & JaSt2-3 & ND$^b$ & -- & -- & -- & -- \\ 
359.8$+$02.4 & & 14.49 & 0.09 & 11.56 & HS$^c$ & HS$^c$  \\ 
359.8$+$03.5 & PHR1731-2709 & 13.84 & 0.09 & 10.93 & 15.5 & 12.4 \\ 
359.8$+$03.7 & & 12.54 & 0.03 & 9.78 & 3.0 & 2.6 \\ 
359.9$-$01.8 & MPA1752-2953 & 13.86 & 0.09 & 12.02 & 24.6 & 20.5 \\ 
359.9$-$02.6 & PHR1756-3019 & 12.90 & 0.03 & 11.25 & 14.2 & 12.6 \\ 
359.9$-$04.5 & M2-27 & 11.12 & 0.02& --  & 3.3 & 3.0 \\ 
359.9$+$01.7 & JaSt2-2 & 14.20 & 0.09& 11.15 & 12.3 & 9.7 \\ 
359.9$+$01.8 & PPA1738-2800 & 14.16 & 0.09 & -- & 7.4 & 4.3 \\ 
\hline
    \end{tabular}
  %\end{minipage}
%\end{table}
%\end{landscape}
%------------------------------------------------------------------------------------------------

%\begin{table}
\caption{$\othreec$ fluxes for 5 PNe just outside our 
region toward the Galactic Bulge.  
All fluxes presented are uncorrected for interstellar absorption. }
\label{tab:extra}
\begin{tabular}{|l l lllll|}
\hline
 & & \multicolumn{3}{|c|}{$-\log$(F[OIII])} &  \multicolumn{2}{|c|}{Diameter ($\arcsec$)}\\ 
PN G name & Usual name & Flux & error & F$_{dered}^d$ & maj & min \\
(1) & (2) & (3) & (4) & (5) & (6) & (7) \\
\hline 
002.4$-$05.0 & PHR1811-2922 & 11.88 & 0.02 & -- & 27.6 & 27.1 \\
003.0$-$05.0 & PHR1812-2848 & 12.14 & 0.02 & -- & 15.3 & 14.1 \\
005.1$+$02.0 & K5-19 & 12.41 & 0.02 & -- & 6.7 & 3.6 \\
305.3$-$03.1 & PHR1315-6555 & 12.35 & 0.02 & -- & 11.4 & 11.4 \\
354.9$-$02.8 & MPA1744-3444 & 13.21 & 0.03 & 11.58 & 9.5 & 8.5 \\
358.7$-$05.1 & SB53 & 11.98 & 0.02 & -- & 14.4 & 11.6 \\
\hline
\end{tabular}
\end{minipage}
\end{table}
\end{landscape}

\clearpage

\label{lastpage}

\end{document}